\documentclass[11pt]{article}
\pdfoutput=1
\usepackage{jcapmod}

\usepackage{booktabs}
\usepackage[english]{babel}
\usepackage{amsmath,amssymb,amsbsy,amstext, amsthm, simplewick}
\usepackage{hyperref}
\usepackage{graphicx}
\usepackage{amsfonts}
\usepackage{amssymb}
\usepackage[small]{caption}
\usepackage{upgreek}
\usepackage[svgnames,dvipsnames,x11names,table]{xcolor}
\usepackage{multirow} 
\usepackage{geometry}
\usepackage[hang,flushmargin]{footmisc}

\usepackage{colortbl}
\definecolor{lightgreen}{cmyk}{0.2, 0, 0.2, 0.2}
\definecolor{lightgray}{cmyk}{0.1,0.2,0,0.1}
\definecolor{lightgray2}{cmyk}{0.1,0.1,0,0.1}

\makeatletter
\newlength{\apb@width}
\newcommand{\autoparbox}[2][c]{\settowidth{\apb@width}{#2}\parbox[#1]{\apb@width}{#2}}

\makeatother

\definecolor{lightgray}{gray}{0.9}

\usepackage[framemethod=default]{mdframed}
\newmdenv[skipabove=7pt,
skipbelow=7pt,
rightline=false,
leftline=false,
topline=false,
bottomline=false,
backgroundcolor=gray!10,
linecolor=gray,
innerleftmargin=5pt,
innerrightmargin=5pt,
innertopmargin=5pt,
innerbottommargin=5pt,
leftmargin=0cm,
rightmargin=0cm,
linewidth=4pt]{eBox}

\numberwithin{equation}{section}

\def\beq{\begin{equation}}
\def\eeq{\end{equation}}

\def\bea{\begin{eqnarray}}
\def\eea{\end{eqnarray}}

\def\d{{\rm d}}

\def\beq{\begin{equation}}
\def\eeq{\end{equation}}
\def\bea{\begin{eqnarray}}
\def\eea{\end{eqnarray}}

\def\d{{\rm d}}

\def\O{{\cal O}}
\def\M{{\cal M}}

\def\Mpl{M_{\rm pl}}

\def\d{{\rm d}}

\def\I{{\cal I}}
\def\K{{\cal K}}

\DeclareRobustCommand{\SkipTocEntry}[4]{}

\definecolor{blue3}{RGB}{31, 119, 180}
\definecolor{red3}{RGB}{	214, 39, 40}
\definecolor{orange3}{RGB}{255, 127, 14}
\definecolor{green3}{RGB}{44, 160, 44}

\begin{document}

\begin{titlepage}
\setcounter{page}{1} \baselineskip=15.5pt 
\thispagestyle{empty}

\begin{center}
{\fontsize{18}{18} \bf Dynamical Constraints on RG Flows and Cosmology}
\end{center}

\vskip 20pt
\begin{center}
\noindent
{\fontsize{12}{18}\selectfont Daniel Baumann,$^1$ Daniel Green,$^2$ and Thomas Hartman$^3$}
\end{center}

\begin{center}
  \vskip 8pt
\textit{$^1$ Institute of Physics, University of Amsterdam, Amsterdam, 1098 XH, The Netherlands}

\vskip 8pt
\textit{$^2$  Department of Physics, University of California, San Diego, La Jolla, CA 92093, USA}

\vskip 8pt
\textit{$^3$  Department of Physics, Cornell University, Ithaca, New York, USA}
\end{center}

\vspace{0.4cm}
 \begin{center}{\bf Abstract}
 \end{center}
 
 \noindent
Sum rules connecting low-energy observables to high-energy physics are an interesting way to probe the  mechanism of inflation and its ultraviolet origin. Unfortunately, such sum rules have proven difficult to study in a cosmological setting. Motivated by this problem, we investigate a precise analogue of inflation in anti-de Sitter spacetime, where it becomes dual to a slow renormalization group flow in the boundary quantum field theory. This dual description provides a firm footing for exploring the constraints of unitarity, analyticity, and causality on the bulk effective field theory. We derive a sum rule that constrains the bulk coupling constants in this theory. In the bulk, the sum rule is related to the speed of radial propagation, while on the boundary, it governs the spreading of nonlocal operators. When the spreading speed approaches the speed of light, the sum rule is saturated, suggesting that the theory becomes free in this limit. We also discuss whether similar results apply to inflation, where an analogous sum rule exists for the propagation speed of inflationary fluctuations.

\end{titlepage}

\restoregeometry

\newpage
\setcounter{tocdepth}{2}
\tableofcontents

\newpage

\section{Introduction}

Effective field theories (EFTs) underlie our understanding of  
virtually every physical phenomenon.   
However, not every effective theory has a short-distance (or high-energy) completion satisfying basic physical principles such as locality and causality~\cite{Adams:2006sv}, and it is important to find precise ways of separating the landscape of consistent low-energy effective theories from the swampland of inconsistent theories~\cite{Vafa:2005ui}.  Identifying the boundaries in the space of EFT parameters can strongly shape our interpretation of current and future measurements of these parameters.  Moreover, the relation between low-energy (IR) observables and the associated high-energy (UV) description can define clear experimental targets or even suggest new types of measurements altogether. 

\vskip 4pt
Cosmology is a domain where understanding the validity of our effective descriptions is particularly crucial. An important feature of inflationary correlation functions is that they probe physics at a fixed energy scale set by the expansion rate during inflation.  We cannot rerun the history of the universe with a larger value of that expansion rate to learn about the UV completion of inflation.  
A central challenge of modern cosmology is to extract the microscopic origin of inflation from the data available, i.e.~from precision measurements of low-energy observables.

\vskip 4pt
A conservative way of assessing the validity of an EFT is by studying the analytic structure of the associated scattering amplitudes~\cite{Adams:2006sv, Distler:2006if, Cheung:2014ega, Baumann:2015nta, Bellazzini:2019xts, Bellazzini:2017fep, Bellazzini:2015cra, Cheung:2016wjt, Cheung:2016yqr, deRham:2018qqo, deRham:2017xox, deRham:2017zjm, deRham:2017imi, deRham:2017avq, Hamada:2018dde, Chen:2019qvr}. This allows certain low-energy parameters to be written as sums over high-energy contributions. These so-called ``sum rules" provide powerful relations between IR observables and UV degrees of freedom. Unitarity often constrains the UV contributions to be positive, leading to interesting positivity constraints on parameters of the EFT.  Unfortunately, these type of arguments are challenging to apply in a cosmological setting, because Lorentz invariance is broken by the time-dependent background. Nevertheless, nontrivial constraints can be derived from causality on subhorizon scales.  These types of constraints, while less stringent than the related sum rules in flat space, hint at deeper structures in the space of EFTs. Since these causality arguments apply equally in a variety of background geometries, they suggest that more rigorous bounds might not be limited by the spacetime geometry in which they were originally derived.  While such results can be compelling and useful, it remains an important challenge to derive these constraints rigorously in cosmological backgrounds (see e.g.~\cite{Baumann:2015nta,Camanho:2014apa, Cordova:2017zej}). 

\vskip 4pt 
Interestingly, inflation has a direct analogue in anti-de Sitter (AdS) spacetime, where it becomes dual to a renormalization group flow in the boundary QFT \cite{Henningson:1998gx,Bianchi:2001de,Bianchi:2001kw,Strominger:2001pn,Strominger:2001gp,McFadden:2009fg}. The problem of finding consistent cosmologies is therefore analogous, though not exactly equivalent, to mapping out the space of possible RG flows. The latter is of course  an important problem in its own right. RG flows between unitary QFTs are highly constrained, most famously by the array of $C$-theorems in two \cite{Zamolodchikov:1986gt}, three \cite{Casini:2012ei},  four \cite{Komargodski:2011vj,Komargodski:2011xv}, and possibly higher (e.g.~\cite{Cordova:2015fha}) dimensions.  These constraints are linked to the role of quantum information in quantum field theory~\cite{Casini:2006es,Casini:2017vbe} and, in holographic theories, the $C$-theorems are realized as constraints on the emergent geometry~\cite{Myers:2010tj}. 
 
\vskip 4pt 
The AdS/CFT dictionary ties the RG flow of a holographic QFT to the breaking of radial translations in the AdS background. The
Goldstone boson associated with this symmetry breaking can be described by an effective theory~\cite{Kaplan:2014dia}. This EFT is in one-to-one correspondence with the EFT of inflation~\cite{Cheung:2007st, Creminelli:2006xe}, after interchanging time in dS with the radial coordinate in AdS.  The problem of finding constraints on the EFT of inflation therefore has an analogous problem in AdS.  Since the boundary dual of the EFT of holographic RG is Lorentz invariant and unitary, the usual constraints on the boundary QFT still apply. 

\vskip 4pt 
In this paper, we will derive new constraints on holographic RG flows and explore their relation to inflationary models. We will study a special class of `slow' holographic RG flows, in which the $C$-function changes slowly along the flow, and the $\beta$-function is constant (though not necessarily small) over a large range of scales. While leading to unconventional RG flows, this is a natural limit to consider because it is analogous to the canonical inflationary evolution, where the Hubble scale changes slowly and the couplings in the EFT of inflation are approximately constant. The constraints on such RG flows, and their derivation, are reminiscent of the four-dimensional $a$-theorem of Komargodski and Schwimmer \cite{Komargodski:2011vj,Komargodski:2011xv}. 
\vskip 4pt
Our main result will be a sum rule related to the speed at which objects fall along the radial direction of AdS, which we will denote by $c_r$. 
This sum rule relates the deviation of this ``infall speed" from the speed of light 
 to an integral of the four-point amplitude of the trace of the stress tensor, $T_\mu^\mu$,  in the dual QFT.  When $c_r$ approaches the speed of light, the sum rule forces an infinite number of operators in the bulk EFT to vanish, suggesting that the theory becomes free in this limit, as conjectured for inflation in~\cite{Baumann:2015nta}.  Our derivation relies on the fact that the EFT of holographic RG flows contains interactions that induce an anomalous dimension for a spin-2 operator in the dual QFT. It is well known that in a CFT, such anomalous dimensions obey positive sum rules~\cite{Nachtmann:1973mr,Komargodski:2012ek,Fitzpatrick:2012yx,Hartman:2015lfa}, and we will show that similar techniques carry over to slow RG flows despite the breaking of conformal invariance.  

\vskip 4pt
On the boundary, the parameter $c_r$ is the speed at which a local wavepacket of $T^{\mu}_{\mu}$ spreads under time evolution, analogous to the butterfly velocity $v_B$ in quantum chaos \cite{Shenker:2013pqa,Mezei:2016wfz}. The operator $T^{\mu}_{\mu}$ is a dynamical probe of the RG flow that explores different scales as a function of time. At early times, the wavepacket
 is localized to a small region, so it is  an operator in 
the UV theory. As the operator spreads,  
it explores larger distance scales. The rate of spreading depends on the details of the RG flow, until eventually it becomes large enough to be an operator in the IR theory. 
Throughout this process, the operator must, of course, spread no faster than the speed of light, so it is natural to find a positive sum rule involving $1-c_r^2$. 

\vskip 4pt
Returning to inflation, the analogue of the infall speed $c_r$ is the propagation speed of inflationary fluctuations, $c_s$, which is often referred to as the inflationary ``speed of sound". This speed is directly related to the leading non-Gaussianities in single-field inflation~\cite{Chen:2006nt}.  As we will show, the same techniques used to derive the sum rule for holographic RG flows can be applied to obtain a sum rule relating inflationary couplings, including $c_s$, to properties of the UV completion. However, in this case, the UV behavior of the corresponding four-point function is poorly understood and there is no known positivity condition like that used to derive the RG constraints. The interpretation of the inflationary sum rule is therefore unclear. If our constraints on $c_r$ can be translated to $c_s$, this would imply that the limit $c_s \approx 1$ is necessarily (canonical) slow-roll inflation with only perturbative higher-derivative interactions. 

\vskip 4pt
While our specific derivation will depend crucially on AdS holography, holography itself demands that our constraint on $c_r$ also has a purely bulk description.  Moreover, one might suspect that the argument in the bulk does not depend on the global properties of the spacetime, as for similar sum rules in pure AdS that match known constraints in Minkowski space. Indeed, we will show that there are equivalent causality constraints for both the EFT of holographic RG flow and the EFT inflation, though they are not precisely equivalent to the constraints implied by our sum rule. These results are suggestive of an inflationary analogue of our constraints, but a rigorous derivation remains an interesting direction for future work.

 \paragraph{Outline} \ \\
The paper is organized as follows: In Section~\ref{sec:Review}, we review the EFT of inflation and its relation to the EFT of holographic RG flows. In Section~\ref{sec:DIS}, we derive a sum rule for the infall speed of the holographic RG flow, using known analyticity properties of deep inelastic scattering. 
In~Section~\ref{sec:dS}, we speculate about the relevance of these results for inflation. We present our conclusions in Section~\ref{sec:conclusions}. 
 Three appendices contain additional material: In Appendix~\ref{app:holographic}, we provide details of the computations described in Section~\ref{sec:DIS}.   In Appendix~\ref{app:betafunction}, we prove that the (indirect) calculation of the anomalous dimension using deep inelastic scattering necessarily agrees with the result of a more direct one-loop calculation.  In Appendix~\ref{app:DBI}, we study the specific effective theory arising from the DBI action and show how, in this case, all EFT parameters are fixed in terms of the propagation speeds $c_s$ and $c_r$.

\paragraph{Notation and conventions} \ \\ 
We will use the following notation for the coordinates on the (A)dS spacetimes: 
\begin{center}
\begin{tabular}{crlll}
\toprule
 \multirow{2}*{$dS_4$}  & bulk & $x^\mu$ &  $\mu=0,1,2, 3$  & $(\nabla \phi)^2 \equiv g^{\mu\nu}\partial_\mu \phi \partial_\nu \phi$\\
& boundary & $x^i$ &  $i=1,2,3$ & $\,(\partial \phi)^2 \equiv \delta^{ij} \partial_i \phi \partial_j \phi$\\
\midrule
 \multirow{2}*{$AdS_{d+1}$}  & bulk & $x^M$ &   $M = 0, \ldots, d$ &  $(\nabla \phi)^2 \equiv g^{M N}\partial_M \phi \partial_N \phi$  \\
& boundary & $x^\alpha$ &  $\alpha = 0,\ldots, d-1$ & $\,(\partial \phi)^2 \equiv \eta^{\alpha\beta} \partial_\alpha \phi \partial_\beta \phi$ \\
\bottomrule
\end{tabular}
\end{center}
We use the first few letters of the greek alphabet ($\alpha, \beta, \ldots$) for the indices of the AdS boundary coordinates to distinguish them from the dS bulk coordinates ($\mu,\nu,\ldots$). 
The dS boundary indices are raised/lowered/contracted with $\delta_{ij}$. The AdS spacetime is in Lorentzian signature, with boundary indices raised/lowered/contracted using $\eta_{\alpha\beta}$ in mostly-plus signature. Bulk and boundary derivatives are denoted by $\nabla$ and $\partial$, respectively. 
We set $\hbar = c = 1$.

\newpage
 \section{From Inflation to RG Flow}
\label{sec:Review}

\begin{figure}[h!]
   \centering
      \includegraphics[scale=1.]{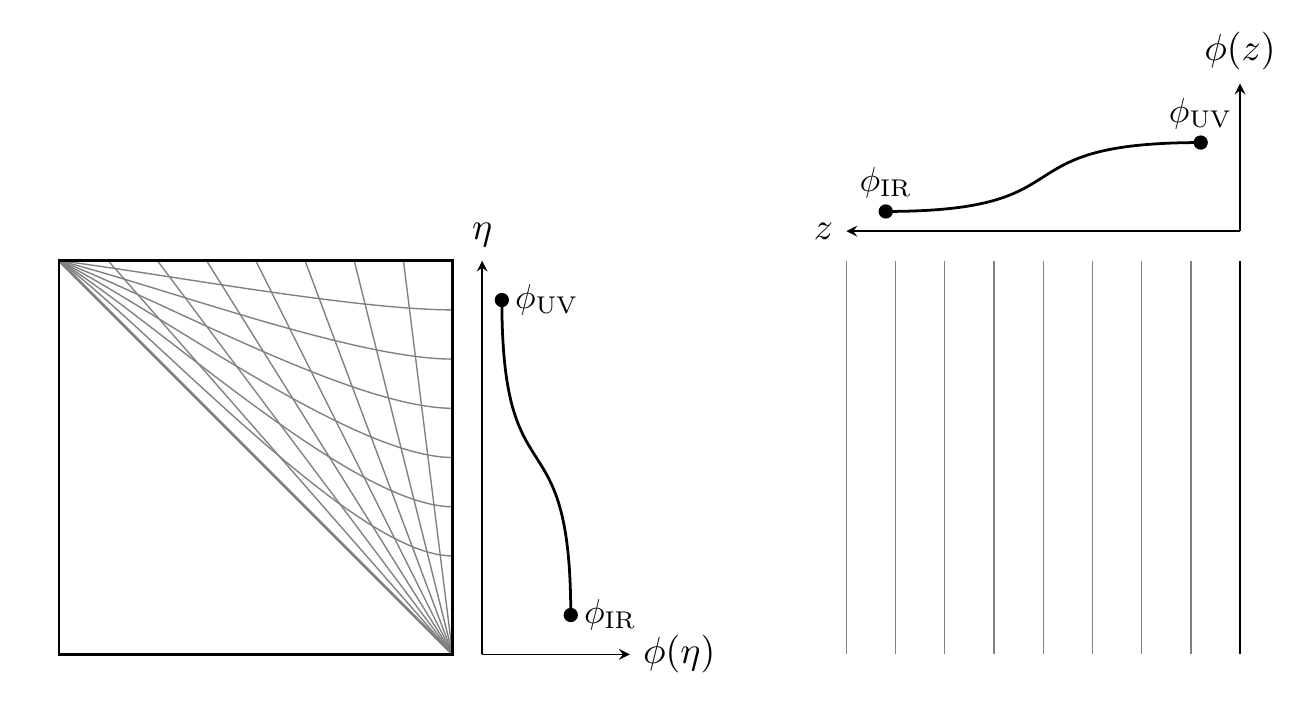} 
      \caption{During inflation ({\it left}), the time dependence of a matter field $\phi(\eta)$ induces a preferred time slicing of the quasi-de Sitter background. 
     The analog in AdS is a slow RG flow ({\it right}) induced by the radial profile $\phi(z)$ of a bulk field. In both cases, the low-energy dynamics can be described by Goldstone fluctuations around the homogeneous background.}
      \label{fig:Penrose}
\end{figure}

Inflation can be understood as the spontaneous breaking of microscopic time translations\footnote{By the breaking of ``time translations" we really mean the breaking of the {\it global part} of time reparameterizations. 
 For a careful discussion of this point, see~\cite{Piazza:2013coa, Baumann:2014nda}.} 
in a quasi-de Sitter spacetime (see Fig.~\ref{fig:Penrose}).  In particular, ending inflation requires a physical clock (often a rolling scalar field) that defines a preferred time-slicing. A powerful way to describe the low-energy dynamics of this symmetry breaking phenomenon is in terms of the Goldstone boson associated with the broken symmetry.
The ``effective field theory of inflation" is the most general effective theory of this Goldstone mode, allowing all terms consistent with the nonlinearly realized symmetry~\cite{Cheung:2007st, Creminelli:2006xe}. From the bottom up, the couplings in this EFT are free parameters, constrained only by naturalness considerations and cosmological observations.
An important question is which Goldstone couplings are allowed from the top down. 
 Are there sum rules and sign constraints imposed by consistency of the UV completion of inflation?
 
 \vskip 4pt 
In this section, we will introduce the analogue of inflation in AdS, where the breaking of radial diffeomorphisms induces a ``slow RG flow" in the boundary field theory (see Fig.~\ref{fig:Penrose}). As for inflation, this RG flow can be described by the Goldstone boson associated with the symmetry breaking~\cite{Kaplan:2014dia}. In Section~\ref{sec:DIS}, we will adopt recent CFT/bootstrap methods to derive new constraints on this RG flow. 

 \subsection{Effective Theory of Inflation}
  \label{sec:EoI}
  
 We begin with a lightning review of the EFT of inflation~\cite{Cheung:2007st, Creminelli:2006xe} (see also~\cite{Baumann:2014nda, Piazza:2013coa}).  Let us write the line element during inflation as
   \beq \label{eq:dsdS}
 \d s^2 = - \d t^2 + a^2(t)\hskip 1pt \d x_i \d x^i \, ,
\eeq
where the Hubble rate $H \equiv \partial_t \ln a$ is approximately, but not exactly, constant.
Fluctuations around this background can be described by the Goldstone boson $\pi(t,x^i)$ associated with the broken time translations.  To construct the Goldstone action, it is useful to define a variable that
transforms linearly under the symmetry:
  \beq
U \equiv t + \pi(t,x^i)\, .
\eeq
Under a diffeomorphism $t \to t + \delta t$, the Goldstone transforms as $\pi \to \pi - \delta t$, so that $U$ transforms as a scalar.
To leading order in derivatives, the most general action of the Goldstone boson is~\cite{Cheung:2007st} 
\begin{align}
S = \int \d t\hskip 1pt \d^{3} x \, a^3\,\bigg[ &\, M_{\rm pl}^2 \left(\frac{1}{2}R  - (3H^2 + \dot H) + \dot H  g^{\mu \nu} \partial_\mu U \partial_\nu U  \right) \nonumber \\
 &\ + \ \sum_{n=2}^\infty \frac{1}{n!} M_n(U) \big[g^{\mu \nu} \partial_\mu U \partial_\nu U + 1\big]^n + \cdots \, \bigg] \, , \label{equ:Action}
\end{align}
where the structure in the first line is enforced by tadpole cancellation.  Taking the so-called {\it decoupling limit}\, ${M_{\rm pl}^2 \to \infty}$ and $\dot H \to 0$, while keeping $M_{\rm pl}^2 \dot H$ fixed, leads to a {\it non-relativistic} QFT in a fixed background~\cite{Baumann:2011su}.  In this limit, dynamical gravity decouples and the Goldstone boson description becomes exact.\footnote{Using the Goldstone boson description while neglecting dynamical gravity is accurate up to  corrections of the order of $H^2 / \Mpl^2$ and $\dot H / H^2$.} In the following, we will assume this limit and set~$M_{\rm pl} \equiv 1$. 

\vskip 4pt
We will take the parameters of the EFT to be time independent. 
This choice is technically natural, since the theory of the fluctuations inherits an  additional global shift symmetry, $U \to U +c$, in this limit.  The resulting action then has an effective time-translation symmetry in the IR, with a corresponding conserved current that can be written as
\beq
t^{\mu} = T^{0 \mu} + j^{\mu}\,,
\eeq 
 where $ T^{0 \mu}$ and $j^\mu$ are the currents associated with time translations in the UV and with the additional shift symmetry, respectively.  An important consequence of this effective time-translation symmetry is the scale invariance of  cosmological correlators.  
 
\vskip 4pt
While scale invariance is an approximate symmetry in all inflationary models,  conformal invariance is generically broken and only re-emerges in slow-roll inflation. To see this, note that Lorentz boosts 
are generated by the currents $K^{\mu \nu \lambda} = x^{\mu} T^{\nu \lambda} - x^{\nu} T^{\mu \lambda}$.  This time the shift symmetry cannot be used to restore the broken boosts, since it does not transform as a tensor.  As a result, the de Sitter isometries, $SO(d,1)$, are broken down to spatial rotations and time translations (i.e.~scale transformations), $SO(d)\times U(1)$. Cosmological correlators then exhibit scale invariance, but are not conformally invariant.
An exception occurs in the slow-roll limit ($M_{n > 1}  \to 0$), where the theory reduces to a free relativistic scalar in the decoupling limit.
Since the theory becomes free, it contains approximately conserved higher-spin currents which allow for an emergent Lorentz symmetry as a diagonal combination of the UV Lorentz group and these global higher-spin symmetries.  
The interactions of slow-roll inflation beyond the decoupling limit are then controlled by the weakly broken conformal symmetry~\cite{Arkani-Hamed:2015bza, Arkani-Hamed:2018kmz, Mata:2012bx}.

\vskip 4pt
Expanding (\ref{equ:Action}) to quadratic order in $\pi$, we find
\begin{align}
{\cal L}_{0} &=   - \dot H \Big[ c_s^{-2} \dot \pi^2 - a^{-2}(\partial \pi)^2\Big] \, , 
\end{align}
where $(\partial \pi)^2 \equiv \delta^{ij} \partial_i \pi \partial_j \pi$ and we have introduced the {\it sound speed} 
\beq
c_s^2 \equiv \frac{ \dot H}{\dot H - 2 M_2}\, ,
\eeq
which describes the speed of propagation of the inflationary fluctuations.
In canonical slow-roll inflation, with $M_2 = 0$, it equals the speed of light, $c_s = 1$, but more generally the broken Lorentz symmetry of the background allows for $c_s \ne 1$.  
Up to quartic order, and  at leading order in derivatives, the Goldstone Lagrangian has the following interactions
\begin{align}
{\cal L}_{\rm int} &=  (1-c_s^{-2})\hskip 1pt \dot H \left[- \dot \pi (\nabla \pi)^2 +  \frac{1}{4}  (\nabla \pi)^4  \right]  - 2 M_3\left[ \frac{2}{3}  \dot{\pi}^3 - \dot \pi^2 (\nabla \pi)^2\right]  + \frac{2}{3}M_4\, \dot \pi^4\, ,\label{Lint}
\end{align}
where $(\nabla \pi)^2 \equiv g^{\mu \nu} \partial_\mu \pi \partial_\nu \pi$.  Additional higher-derivative operators should also be included in the EFT~\cite{Cheung:2007st}, but they  often give suppressed contributions to cosmological correlators.

\vskip 4pt
Because Lorentz symmetry is broken, time and space derivatives cannot be treated on an equal footing. 
To avoid this complication, we rescale the spatial coordinates as $x \to \tilde x = c_s^{-1} x$.
This rescaling restores a fake Lorentz symmetry in the quadratic Lagrangian
\beq
\tilde {\cal L}_{0} = c_s^3 {\cal L}_0 =  \frac{1}{2} f_\pi^4 \Big[ \dot \pi^2 - a^{-2}(\tilde \partial \pi)^2\Big] \equiv - \frac{1}{2}  (\tilde \nabla \pi_c)^2\, ,  \label{equ:S2x}
\eeq
where $f_\pi^4 = 2 |\dot H| c_s$ is the symmetry breaking scale and $\pi_c \equiv f_\pi^2 \pi$ is the canonically-normalized Goldstone mode. The dimensionless power spectrum of primordial curvature perturbations, $\zeta=-H\pi$, is given by  $2\pi \Delta_\zeta = (H/f_\pi)^2$, and the observed size of $\zeta$-fluctuations, $\Delta_\zeta \approx 4.5 \times 10^{-5}$, implies the hierarchy $f_\pi = 59 \hskip 1pt H$.  In terms of the rescaled coordinates, time and space are again on an equal footing, at least perturbatively, and 
the  relative impact of each operator can be read off from the couplings alone.  

\vskip 4pt
It is useful to measure the parameters of the EFT of inflation in terms of the symmetry breaking scale:
\beq\label{eq:Mncs}
M_n \equiv c_n \frac{f_\pi^4}{c_s^{2n-1}}\, ,
\eeq
where $c_2 \equiv \frac{1}{4}(1-c_s^2)$ and the factors of $c_s$ were introduced to ensure that $c_n \sim O(1)$ is natural 
even for small values of $c_s$~\cite{Baumann:2015nta}.  In DBI inflation (see Appendix~\ref{app:DBI}), all EFT parameters $c_n$ are fixed in terms of the sound speed $c_s$ (or equivalently $c_2$); in particular, $c_3 = -6 \hskip 1pt c_2^2$ and $c_4 = 60 \hskip 1pt c_2^3$.
The interaction Lagrangian (\ref{Lint}) then becomes 
\begin{align}
\label{equ:Lint3X}
\tilde{\cal L}_{\rm int} &=  -\frac{2}{\Lambda^{2}} \left[ c_2\, \dot \pi_c (\tilde \nabla \pi_c)^2 + \left((1-c_s^2) \hskip 1pt c_2 + \frac{2}{3}c_3\right) \dot \pi_c^3 \right] \\
 &\ + \frac{1}{2 \Lambda^{4}} \left[c_2\, (\tilde \nabla \pi_c)^4 + 2\Big((1-c_s^2)  \hskip 1ptc_2 + 2c_3\Big) \dot \pi_c^2 (\tilde \nabla \pi_c)^2 +\left((1-c_s^2)^2 \,c_2 + 4 (1-c_s^2)c_3+ \frac{4}{3}c_4\right) \dot \pi_c^4\right] , \nonumber
\end{align}
where $\Lambda \equiv f_\pi c_s$.
We see that a small sound speed, $c_s \ll 1$, implies large interactions because of the reduced cutoff scale $\Lambda$ in (\ref{equ:Lint3X}). 
The corresponding non-Gaussianity of the initial conditions is~\cite{Baumann:2014nda} 
\beq
f_{\rm NL} \approx \left[ - \frac{85}{81}\, c_2 + \frac{40}{243} \,c_3 \right]   \left(\frac{f_\pi}{\Lambda}\right)^2\,  ,
\eeq
where $f_{\rm NL} \Delta_\zeta$ measures the deviation from Gaussianity.  
Constraints on the CMB bispectrum~\cite{Akrami:2019izv} translate into $f_{\rm NL} =  -26 \pm 47$ (68\%\,CL), which  means that the universe is Gaussian at the $10^{-3}$ level.  Marginalizing over $c_3$, this implies the following bound on the inflationary sound speed
\beq
c_s > 0.021 \, \quad (95 \%\,{\rm CL})  \,.
\eeq
For comparison, canonical single-field slow-roll inflation requires $c_s > 0.34$~\cite{Baumann:2015nta} to be consistent with a single weakly coupled scalar at the scale $f_\pi$. Deviations from $c_s=1$ in that case only arise from perturbative higher-derivative interactions, such as $(\nabla \phi)^4$. 

 \subsection{EFT of Holographic RG Flows}
 \label{sec:EFTofRG}

 A holographic RG flow is a deformation of AdS/CFT in which a relevant operator is added to the boundary action,
 \beq\label{equ:deform}
 S = S_{\rm CFT} + \lambda \int \d^d x\, \Phi(x) \, .
 \eeq
The scalar operator $\Phi$ has scaling dimension $\Delta < d$, which breaks conformal invariance and triggers an RG flow to the infrared. The trace of the stress tensor, $T \equiv T_\mu^\mu$, is related to the operator $\Phi$ via the beta function,  $T  = \beta \hskip 2pt \Phi$.
Poincar\'{e} invariance in preserved. The geometry dual to the perturbed field theory is
 \beq
 \d s^2 =  \d r^2 + a^2(r)\hskip 1pt \d x_\alpha \d x^\alpha \, ,
\eeq 
where the scale factor asymptotically takes the AdS form $a(r) \to e^{ r/\ell}$ as $r \to \infty$, but differs in the interior. We take the boundary metric to be $d$-dimensional Minkowski spacetime. The solution is supported by nontrivial matter fields. We focus on single-field flows, in which the only fields that participate are the metric and the bulk scalar $\phi$ dual to the operator $\Phi$, with background profile $\phi = \phi_0(r)$. According to the usual AdS/CFT dictionary, the asymptotic behavior of $\phi_0(r)$ is related to the QFT coupling constant $\lambda$.

\vskip 4pt
The analogue of inflation is a quasi-AdS RG flow, where the ``Hubble parameter" $H \equiv \partial_r \ln a$ is approximately, but not exactly, constant. This is equivalent to the assumption that the holographic $C$-function~\cite{Myers:2010tj}, 
\beq
C(r) = \frac{\pi^{d/2}}{\Gamma[d/2] } \left(\frac{M_{\rm pl}}{H}\right)^{d-1} \, ,
\eeq
is nearly constant. Fluctuations around the ``slow-flow" background can be organized into an effective field theory \cite{Kaplan:2014dia}. In fact, it is easy to see that
there is a one-to-one map between the EFT of inflation described above and the EFT of holographic RG flow. 
As before, we introduce the Goldstone boson via $U \equiv r + \pi(r,x^\alpha)$ and write
its effective action as 
\begin{align}
S \,\supset\,   \int \d r\hskip 1pt \d^{d} x \, \sqrt{g}\, \left[ H'  (\nabla U)^2 + \sum_{n=2}^\infty \frac{1}{n!} M_n(U) \left[(\nabla U)^2 - 1\right]^n + \cdots  \right]  ,
\label{equ:AdSAction}
\end{align}
where $(\nabla U)^2 \equiv g^{MN} \partial_M U \partial_N U$. 
Like in the EFT of inflation, we will take the parameters of the Goldstone action to be nearly constant, in the sense that $\partial_r M_n / M_n \ll H$.  This choice is again protected by the shift symmetry\footnote{Global symmetries in the bulk act on the boundary QFT, but do not have an associated conserved current, which would require an additional gauge field in the bulk.  This  is particularly unusual for a shift symmetry, since $\phi \to \phi+c$ implies the existence of an exactly marginal operator, such that every CFT along the conformal manifold is equivalent, i.e.~shifting the coupling by a constant returns the same CFT.  See \cite{Harlow:2018jwu} for a recent discussion.}
$U \to U + c$.

\vskip 4pt
At quadratic order in $\pi$, we get
\beq
{\cal L}_{0} =  H' \Big[   c_r^2 (\pi')^2 + a^{-2}(\partial \pi)^2\Big]\,  ,
\eeq
where we have introduced\hskip 1pt\footnote{Note that our definition of the propagation speed is different from that used in~\cite{Kaplan:2014dia}.} 
\beq\label{eq:cr}
c_r^2 \equiv \frac{H' + 2M_2}{ H' }\, .
\eeq
Note that $H' < 0$, so that $c_r <1$ requires $M_2 > 0$. In the bulk, the parameter $c_r$ is the speed at which particles fall into the AdS spacetime.  We will therefore call it the {\it infall speed}.
In the dual boundary QFT, the parameter $c_r$ is the speed at which certain smeared operators spread as a function of time. We will discuss this further in \S\ref{sec:spreading}.

\vskip 4pt
Up to quartic order, the Goldstone interactions are
\beq
{\cal L}_{\rm int} =  -(1-c_r^2)\hskip 1pt H' \left[ \pi' (\nabla \pi)^2 +  \frac{1}{4}  (\nabla\pi)^4  \right]  + 2 M_3\left[ \frac{2}{3}  (\pi')^3 + (\pi')^2 (\nabla \pi)^2\right]  + \frac{2}{3}M_4 (\pi')^4\, .\label{Lint2}
\eeq
Defining the rescaled radial coordinate $\tilde r = c_r^{-1} r$, a fake AdS invariance can be restored in the quadratic Lagrangian \beq\label{eq:quadraticpi}
\tilde{\cal L}_0 = c_r {\cal L}_0 = - \frac{1}{2}f_\pi^{d+1} \Big[ (\partial_{\tilde r}\pi)^2 + a^{-2} (\partial \pi)^2\Big]  \equiv - \frac{1}{2}  (\tilde \nabla \pi_c)^2 \, ,
\eeq
where $f_\pi^{d+1} \equiv 2|H'|c_r$ is the symmetry breaking scale and $\pi_c^2 \equiv f_\pi^{d+1} \hskip 1pt \pi^2$ is the canonically-normalized Goldstone mode. 
After the rescaling, the scale factor is 
$a(\tilde r )\approx  e^{\tilde H \tilde r}$,
where  
\beq
\tilde H \equiv c_r H\, .
\eeq
The interaction Lagrangian (\ref{Lint2}) becomes 
\begin{align}
\tilde{\cal L}_{\rm int} = &\ \frac{2}{f_\pi^{(d+1)/2}} \left[ \frac{c_2}{c_r} a^{-2} \partial_{\tilde r} \pi_c (\partial \pi_c)^2  + \left(\frac{c_2 + \frac{2}{3}c_3}{c_r^3}\right) (\partial_{\tilde r} \pi_c)^3 \right] \label{equ:Lint3} \\
 &+ \frac{1}{2 f_\pi^{d+1}} \left[c_2\, a^{-4} (\partial \pi_c)^4+ \frac{2 (c_2 + 2c_3)}{c_r^2} a^{-2} (\partial_{\tilde r} \pi_c)^2 (\partial \pi_c)^2  +\left(\frac{c_2 + 4c_3+ \frac{4}{3}c_4}{c_r^4}\right) (\partial_{\tilde r} \pi_c)^4 \right]    ,  \nonumber
\end{align}
where we have defined
\beq\label{eq:Mn}
M_n \equiv f_\pi^{d+1}\, \frac{c_n}{c_r} = -2 H' \, c_n \,.
 \eeq 
Notice that the scaling with $c_r$ in (\ref{eq:Mn}) is different from the scaling with $c_s$ in (\ref{eq:Mncs}).  
With this choice of scalings, we have  $c_n \sim {\cal O}(1)$ in both cases. In Appendix~\ref{app:DBI}, we demonstrate this for the DBI action.

\vskip 4pt
A crucial feature of the Goldstone Lagrangian is that the infall speed $c_r$ controls not only the kinetic term, but also the $(\partial \pi)^4$ interaction, according to the relation 
\beq\label{c2cr}
c_2 = \frac{1}{4}(1 - c_r^2) \, .
\eeq
Note that $c_r$ measures the propagation speed in the radial direction of AdS, while fluctuations in the transverse directions always move at the speed of light.  However,  the $(\partial \pi)^4$ interaction does affects the transverse speed of propagation if we deform the background by a gradient in the transverse directions~\cite{Adams:2006sv}. The relation \eqref{c2cr} therefore connects the propagation speed in the radial direction to the transverse propagation speed around nontrivial backgrounds. 

\subsection{Scale-But-Not-Conformal QFT}
\label{sec:dual}

Inflationary correlators typically exhibit approximate scale invariance, consistent with cosmological observations.  On the other hand, the inflationary background generically breaks conformal invariance.  By construction, the bulk description of slow RG flows exhibits the same phenomenon.  It is natural to wonder how this manifests itself on the boundary QFT, especially because scale-but-not-conformal invariance is highly constrained in relativistic QFTs~\cite{Luty:2012ww,Dymarsky:2013pqa,Dymarsky:2014zja}.

\vskip 4pt
The breaking of conformal invariance on the boundary is captured by the trace of the stress tensor $T$. To normalize the two-point function of $T$, we recall that it is dual to the curvature perturbation in the bulk, $\zeta \approx -H\pi = - \tilde{H} \pi/c_r$. 
Using~(\ref{eq:quadraticpi}), we then have 
\beq
\langle T(\vec{k}) T(-\vec{k}) \rangle'  = \beta^2 \langle \Phi(\vec k) \Phi(-\vec{k}) \rangle' \approx c_r^2 \left(\frac{f_\pi}{\tilde H} \right)^{d+1} k^d \, ,
\eeq
for odd dimensions $d$. Letting $\langle \Phi(\vec k) \Phi(-\vec{k}) \rangle' \equiv k^d$, this implies
\beq
|\beta| = c_r\hskip 1pt  \left(\frac{f_\pi}{\tilde H} \right)^{(d+1)/2} \gg c_r^{-2} \geq 1 \, ,
\label{equ:beta}
\eeq
where we imposed  $f_{\pi}^{(d+1)/2} c_r^3 \gg \tilde{H}^{(d+1)/2}$, in order for the bulk EFT \eqref{equ:Lint3} to be weakly coupled and under control.
In addition, the decoupling of gravity in the bulk corresponds to taking $N \to \infty$ on the boundary. In the boundary QFT, we are therefore working in the limit
\beq\label{equ:bnlim}
N^2 \gg |\beta| \gg 1 \ .
\eeq
As we will now discuss, the fluctuations are still scale-invariant, despite the large $\beta$-function.

\vskip 4pt
For concreteness, let us consider a slow RG flow that is triggered by turning on a slightly relevant operator with a large coupling. Specifically, we take the operator $\Phi$ in \eqref{equ:deform} to have dimension $\Delta = d-\epsilon$. The decoupling limit in the bulk then corresponds to taking the coefficient $\lambda \to \infty$, while holding $\beta = - \lambda \epsilon \gg 1$ fixed and much smaller than $N^2$.  In this limit, the  boundary correlators become exactly scale invariant, 
i.e.~although $|\beta| \gg 1$, scale invariance is only broken at ${\cal O}(\epsilon)$. 
Conformal invariance, on the other hand, can be broken by much larger amounts. This does not violate existing constraints on 
scale-but-not-conformal theories (e.g.~\cite{Luty:2012ww,Dymarsky:2013pqa,Dymarsky:2014zja}), because of the unusual limit that is being taken.\footnote{It remains a possibility that even the behavior in the decoupling limit is inconsistent. 
After all, scale invariance of the boundary correlators originates from a global shift symmetry in the bulk, which is generally thought to be forbidden in a theory of quantum gravity.  We are assuming that the breaking of this symmetry decouples as $\Mpl \to\infty$, but it is conceivable that there is some fundamental obstacle to taking the limit in this way.}  Had we fixed $\lambda$, while taking the limit $\epsilon \to 0$, we would have recovered scale and conformal invariance as expected.  What is more remarkable is that in the decoupling limit scale invariance and conformal invariance are broken\footnote{Given that the metric is still approximately AdS, the graviton and any spectator fields are approximately scale and conformally invariant.  In this sense, the unusual breaking pattern is isolated to a sub-sector of the full QFT. } at different orders of~$\lambda$.

\subsection{Operator Growth in RG Flows}
\label{sec:spreading}

\begin{figure}[h!]
\begin{center}
\includegraphics[width=1\textwidth]{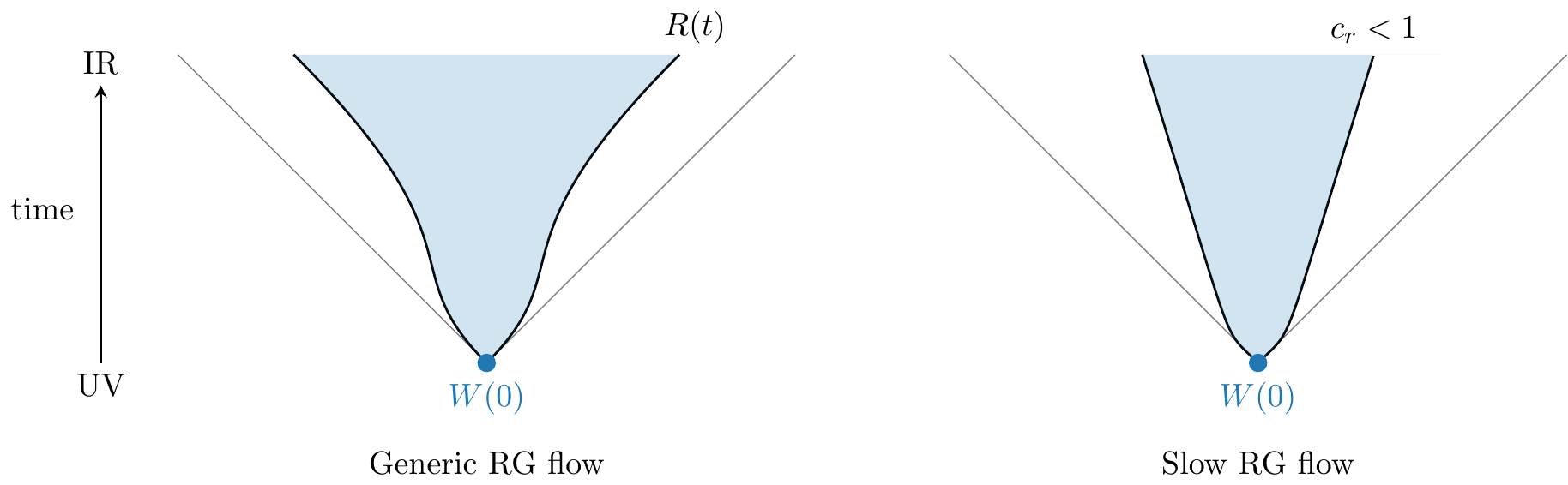}
\end{center}
\vspace{-0.35cm}
\caption{Illustration of the growth of an operator in the vacuum state of a QFT for a generic RG flow~({\it left}) and the slow RG flow ({\it right}). The operator $W$ is smeared within a small region near the origin. As the operator spreads in space, it probes larger scales, and effectively transitions from the UV CFT at early times to the IR CFT at late times. The growth curve $R(t)$ depends on the UV operator and the details of the RG flow. In the slow RG limit, the operator grows linearly over a large range of scales. The speed of operator growth $\d R/dt$ is then equal to the AdS infall speed $c_r$.\label{fig:growthcurves}} 
\end{figure}

The infall speed $c_r$ introduced in \S\ref{sec:EFTofRG} has an interesting interpretation in the dual QFT on the boundary: it is the speed at which certain non-local operators spread in space, as a function of time (see Fig.~\ref{fig:growthcurves}). It is also a probe of the physics along the RG flow.  As we will now explain, this connection between operator growth and RG flow is relevant also in non-holographic QFTs.

\vskip 4pt
A relativistic QFT flows to scale-invariant fixed points in the UV and IR. It is widely believed, and in many cases proven, that the fixed points are also conformally invariant \cite{Nakayama:2013is}.  
Typically, the RG flow between the fixed points is viewed as an abstract flow in the space of Lagrangians, or as a change in the effective description of a given theory as a function of scale. It can also be probed dynamically. To illustrate the latter point of view, let us consider the thought experiment pictured in Fig.~\ref{fig:growthcurves}.   An operator $W$ is localized in a small region of size $R(0) \ll M^{-1}$ near the origin, where $M$ is a characteristic mass scale of the QFT. 
Effectively, $W$ is an operator in the UV CFT. Under time evolution, the Heisenberg operator $W(t) = e^{iHt}W e^{-iHt}$ spreads out in space, exploring different scales along the RG flow. Eventually, when it becomes of size $R(t) \gg M^{-1}$, it is an operator in the IR CFT.   
The growth curve $R(t)$ depends on the choice of UV operator $W$ and on the details of the RG flow.\footnote{As a simple example, we could start the experiment by inserting a local operator $W = {\cal O}(0)$ in an undeformed CFT. The operator $W$ then spreads at the speed of light, so the growth curve $R(t)$ is just the usual lightcone. More general operators, or even local operators in non-conformal theories, may not spread at the speed of light, so $R(t)$ can be nontrivial.} It should be possible to define $R(t)$ in general, but we will only give a precise definition  in holographic theories, where this experiment is interpreted as dropping a particle into the bulk and letting it fall in the radial direction. 

\vskip 4pt
To interpret the infall speed $c_r$, we choose the operator $W$ to be a wavepacket localized near the origin, with timelike momentum, built by smearing the trace of the stress tensor:
\beq
W \equiv \int \d^d x \,  e^{-i\omega t} e^{-(t^2 + \vec{x}^{\hskip 1pt 2})/\sigma^2} \,T_\mu^\mu (t,\vec{x}\hskip 1pt) \, .
\eeq
The width and frequency of the wavepacket satisfy $\sigma  M \ll 1$ and  $\omega \sigma \gg 1$, so that at early times, the evolution of $W$ is controlled by the UV CFT. For $t \ll M^{-1}$, the growth is linear at the speed of light. 
In the bulk theory, $W$ creates a $\pi$ particle that starts near the conformal boundary at $x=z=0$, and then falls radially inward (see for example \cite{Afkhami-Jeddi:2017rmx}). The purpose of the wavepacket is to localize this particle on a worldline, rather than sending a wave of $\pi$ in all directions into the bulk. 

\vskip 4pt
We can now give the precise definition of the operator growth curve $R(t)$: It is the size of the boundary region necessary to reconstruct the infalling $\pi$-particle at time $t$.
 In empty AdS, a radially falling particle would travel on a null geodesic, with ${dr/dt} = -e^{r/\ell}$. At time~$t$, it can be reconstructed on the boundary in a region of size $R(t) = t$, so the dual operator on the boundary spreads at the speed of light. This is a special feature of a CFT, and once conformal invariance is broken, we should not expect this operator to spread at the speed of light. Indeed, in a background corresponding to a  nontrivial RG flow, it spreads more slowly. The full curve $R(t)$ can be calculated by entanglement wedge reconstruction, along the lines of \cite{Mezei:2016wfz}. It is the size of the boundary region whose holographic entanglement surface reaches to the point $r(t)$ in the bulk. There are generally three effects from the RG flow that need to be accounted for: the change in radial null geodesics, the change in the shape of the entanglement wedge, and the  interactions of the field $\pi$ with the background $\phi_0(r)$.
Having said that, we will not need this full machinery, because in the slow-RG limit, the metric is very close to AdS. Only the last effect, self interactions, modifies the infall speed of the Goldstone particle, $c_r = -a^{-1} dr/dt$. The reconstruction region on the boundary is simply $R = a^{-1}$, as in empty AdS. The rate of operator spreading is therefore equal to the infall speed:
\beq
\frac{dR}{dt} = c_r \, .
\eeq
Another assumption of the slow-RG limit is that $c_r$ is constant over a large range of scales. This leads to linear operator growth, inside an effective lightcone of slope $c_r^{-1}$, as in the right panel of Fig.~\ref{fig:growthcurves}.

\vskip 4pt
There is an interesting parallel between this picture and the spread of quantum chaos at finite temperature \cite{Shenker:2013pqa,Roberts:2014isa}. Chaos spreads linearly at the butterfly velocity $v_B$. In holographic theories, the butterfly velocity can also be calculated from the speed of radial infall, but the experiment is done in a black hole geometry rather than an RG flow \cite{Mezei:2016wfz,Mezei:2016zxg}. The picture of operator growth that we have just described is very similar to the spread of chaos, but it is in the vacuum state of a non-conformal QFT, rather than a thermal state, and the butterfly velocity~$v_B$ is replaced by the RG velocity $c_r$. It would be interesting to explore this parallel more deeply.

\section{Bounds from Deep Inelastic Scattering}
\label{sec:DIS}

Deep inelastic scattering (DIS) is an interesting way to study strongly coupled theories  (see Fig.~\ref{fig:DIS}).  While the target may not have a small parameter in which to explore the system, DIS introduces an additional weakly coupled probe that provides a new degree of control.  Not only has DIS proved essential for the discovery of asymptotic freedom in QCD, but it has also been a valuable tool for understanding the structure of more general QFTs (including CFTs).  Moreover, the DIS amplitude is also a natural object to compute using holography, as it is essentially a bulk four-point function in momentum space.  In this section, we will explore how constraints on the DIS amplitude in the boundary QFT will lead to sum rules for bulk couplings. 
 
\vskip 4pt
We begin, in \S\ref{sec:review}, with a brief review of deep inelastic scattering and derive a sum rule for the spin-$s$ moments of the DIS amplitude. In~\S\ref{sec:holo_sum}, we specialize to QFTs with holographic duals.   
We first relate the DIS amplitude to a four-point function of a scalar operator $\O$ and then show that the spin-2 moment of the amplitude is related to the spin-2 anomalous dimension of $\O$  \cite{Nachtmann:1973mr,Komargodski:2012ek}.
In \S\ref{equ:matching}, we compute this anomalous dimension holographically in terms of the bulk interactions of the Goldstone EFT.  There are various approaches in the literature to calculate these anomalous dimensions holographically \cite{Heemskerk:2009pn,Fitzpatrick:2010zm,Hijano:2015zsa}, but they typically rely on conformal invariance. Instead, we will calculate the DIS amplitude directly using momentum-space Witten diagrams, and extract the anomalous dimension from the result. We then present a sum rule for the EFT parameters $c_r$ and~$c_3$. Finally, we discuss the theoretical implications of our result, first for $c_r \approx 1$ (\S\ref{sec:implications}) and then for $c_r \ll1$ (\S\ref{sec:smallcr}).

\subsection{Review of Deep Inelastic Scattering}
\label{sec:review}

In any DIS calculation, the probe couples to the QFT of interest through a specific operator, which, in the context of QCD, is usually the electromagnetic current.  We will denote the probe particle by $\phi$  and couple it to a scalar operator $\O$  via the interaction $g \hskip 1pt \phi \hskip 1pt\O$, where the coupling $g$ can be chosen to be arbitrarily weak.  Our analysis will hold for any scalar operator $\O$, but we will eventually specialize to the case $\O = T$, where $T$ is the trace of the stress tensor.  

\begin{figure}[t!]
   \centering
      \includegraphics[scale=0.8]{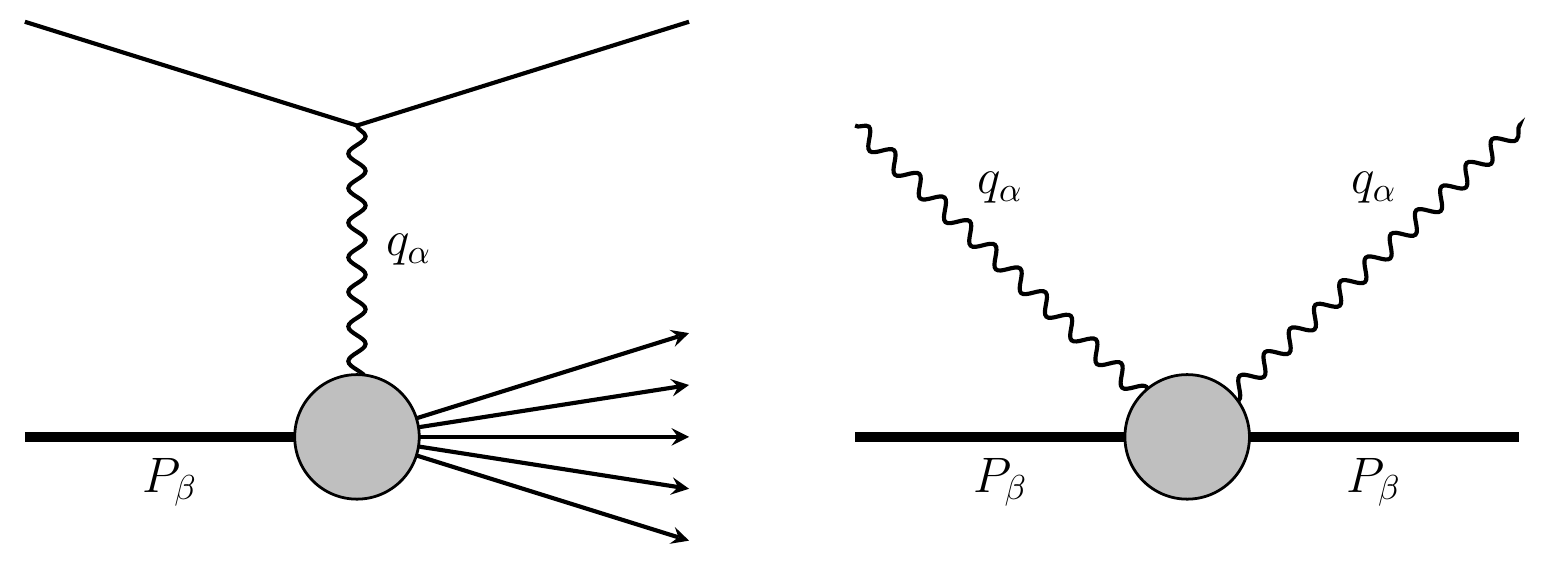} 
      \caption{Illustration of deep inelastic scattering ({\it left}) and the DIS amplitude ({\it right}).}
      \label{fig:DIS}
\end{figure}

\vskip 4pt
The probe particle scatters off a state in the strongly coupled theory by exchanging ${\cal O}$ with momentum $q$
 (see Fig.~\ref{fig:DIS}).  The amplitude for the elastic scattering $\phi P \to \phi P$ is related by unitarity to the inelastic cross section via
\beq
2\hskip 1pt{\rm Im} \hskip 1pt\M(\phi P \to \phi P) = \sum_f \int \d \Pi_f\, |\M(\phi P \to f )|^2\, ,
\eeq
where $f$ is any possible final state of phase space density $\d \Pi_f$. At leading order in $g$, these amplitudes are given by 
\begin{align}
i\M(\phi P \to f ) &= i g \int \d^d y\, e^{i q \cdot y} \langle f | \O(y) | P\rangle\, , \\
i \M(\phi P \to \phi P)&= (i g)^2 \int \d^d y \, e^{i q\cdot y} \,\langle P | \,{\mathbb T}(\O(y) \O(0) ) | P \rangle \, , \label{equ:DIS0}
\end{align}
where ${\mathbb T}$ denotes time ordering.
The elastic amplitude is a particularly useful object for two reasons: {\it i}\hskip 1pt) at sufficiently low center-of-mass energies, the scattering is controlled by the operator product expansion (OPE) and {\it ii}\hskip 1pt) the imaginary part is related to the total inelastic cross section and is necessarily positive.  Combining these insights with analyticity of the amplitudes has led to interesting constraints on the low-energy description, including sign and convexity constraints on anomalous dimensions of minimum-twist operators~\cite{Nachtmann:1973mr,Komargodski:2012ek,Komargodski:2016gci}.

\vskip 4pt
The use of the DIS amplitude is natural in holographic theories, because bulk interactions control the anomalous dimensions and OPE coefficients in the boundary QFT~\cite{Heemskerk:2009pn,Fitzpatrick:2010zm}.  When the bulk theory is Lorentz invariant and the boundary theory is a CFT, sign constraints on the anomalous dimensions of boundary operators~\cite{Hartman:2015lfa} reproduce known sign constraints on the bulk EFT, derived from the positivity of scattering amplitudes~\cite{Adams:2006sv}.  When the bulk breaks AdS invariance, the bulk scattering methods are less effective~\cite{Baumann:2015nta}, but we still  expect the bounds from the DIS amplitudes on the boundary to apply.  

\vskip 4pt
In the following, we will study the DIS amplitude defined in (\ref{equ:DIS0}): 
\beq
A(q,P) \equiv i \int \d^d y \, e^{i q\cdot y} \,\langle P | \,{\mathbb T}(\O(y) \O(0) ) | P \rangle\, ,
\label{equ:DIS}
\eeq
which, in the limit of large $q^2$, can be simplified by using the OPE.
  For real operators $\O$, the OPE only gets contributions from even spins. Moreover, only ``primary" operators\hskip 1pt\footnote{For the lack of a better term, we call operators that cannot be expressed as derivatives of other operators ``primary", in analogy with primaries in a CFT.  However, it is important that our analysis does not assume conformal invariance at any point.} need to be considered, since operators that can be expressed as a total derivative do not contribute to the relevant correlation function because of momentum conservation.  We can therefore write
  \beq
\O(y) \O(0)  \sim \sum_{s=0,2,\cdots} \sum_{n \in {\cal I}_s} f_s^{(n)}(y)\, y^{\alpha_1} \cdots y^{\alpha_s}\, \O^{(n)}_{\alpha_1 \ldots \alpha_s}(0)\, , \label{equ:OPE}
\eeq
where ${\cal I}_s$ denotes the set of primary operators of spin $s$.  
Imposing scale invariance, the coefficient functions in (\ref{equ:OPE}) are constrained to take the form 
\beq
f_s^{(n)}(y) \equiv c_{s}^{(n)}\, y^{\tau_s^{(n)} - 2 \Delta} \, ,
\eeq
where  $\tau_s^{(n)} \equiv \Delta_s^{(n)} -s $ is the twist of the operator. 
Using
\beq
\langle P | \O^{(n)}_{\alpha_1 \ldots\alpha_s}(0) | P \rangle = d_{s}^{(n)} [ P_{\alpha_1} \ldots P_{\alpha_s} - {\rm traces}] \, ,
\eeq
where $d_{s}^{(n)}$ are dimensionful coefficients,
 we can write the DIS amplitude as
 \beq
 A(q,P)=  \sum_{s=0,2,\cdots} \sum_{n} \tilde{c}_{s}^{(n)} d_{s}^{(n)}\Big( [P\cdot \partial_q]^s - \text{traces} \Big)  \, q^{2 \Delta - \tau_s^{(n)} - d} \, ,
 \eeq
 with the momentum-space OPE coefficients
 \beq\label{momope}
 \tilde{c}_s^{(n)}  =2^{d + \tau_s^{(n)} - 2\Delta}\pi^{d/2} \frac{\Gamma(d/2 - \Delta + \tau_s^{(n)}/2)}{\Gamma(\Delta - \tau_s^{(n)}/2)}\,  c_s^{(n)}  \, .
 \eeq
It is convenient to switch to an alternative parameterization in terms of   
\beq
x \equiv \frac{q^2}{2 P \cdot q}\, .
\eeq
High-energy scattering then corresponds to the limit $x\to 0$. Keeping only the leading terms for large~$q^2$, 
we find
 \beq
  A(x,q^2 ) \approx   \sum_{s=0,2,\cdots}  \tilde{c}_s^{(*)} d_{s}^{(*)} \, x^{-s} (q^2)^{\Delta  -d/2 - \tau_s^{(*)} /2} \, , \label{equ:DIS-OPE}
 \eeq
 where the asterisks denote quantities associated with minimal-twist operators. 
 
    \begin{figure}[t!]
   \centering
      \includegraphics[scale=1.]{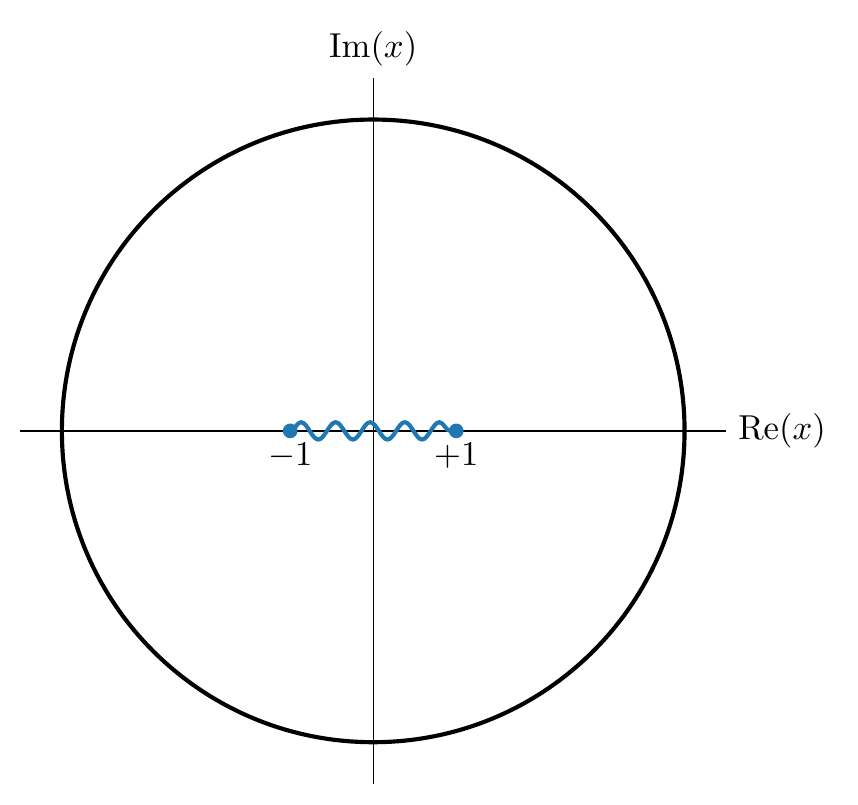} 
      \caption{Illustration of the analytic structure of the DIS amplitude.}
      \label{fig:contour}
\end{figure}

 \vskip 4pt
We have just seen that the OPE is useful for large $x$, while the physically relevant regime is $|x| \in [0,1]$. The OPE limit and the small-$x$ regime can be related by a contour deformation in the complex $x$-plane. Note that kinematics requires $-(P+q)^2 \geq 0$  
 and the DIS amplitude has a branch cut $x \in [-x_c,x_c]$, with $x_c = (1 + {P^2/q^2})^{-1}$.  Since we work at large $q^2$ and fixed $P^2$, we can set $x_c = 1$.  Let us consider the  $s$-th moment of the amplitude, $a_s(q^2)$, defined as the integral of $x^{s-1} A(x,q^2)$ around the closed contour shown in Fig.~\ref{fig:contour}. This moment is determined by the discontinuity across the branch cut: 
\beq
a_s(q^2) \equiv \frac{1}{2 \pi i }\oint \d x\, x^{s-1} A(x,q^2) =   \frac{2}{\pi} \int_{0}^{1} \d x\, x^{s-1}\, {\rm Im}[A(x,q^2)] \, .
\label{equ:moment}
\eeq
To write this sum rule, we have assumed polynomial boundedness of the amplitude for small values of $x$, i.e.~$\lim_{x \to 0} A(x,q^2) \leq x^{-M+1}$,
for some integer $M$. (Below, we will assume that the bound is strong enough to permit a spin-2 sum rule.)
For large $q^2$, the DIS amplitude (\ref{equ:DIS-OPE}) then implies 
\beq
\begin{aligned}
\hspace{1.1cm}a_s(q^2) &\ \to\   \tilde{c}_s^{(*)} d_{s}^{(*)}\, (q^2)^{\Delta-d/2 - \tau_s^{(*)} /2} \\[4pt] &\ =\  \frac{2}{\pi}\int_0^1 \d x\, x^{s-1}\, {\rm Im}[A(x, q^2\to \infty)] \, ,
\end{aligned}
\label{equ:SUM0}
\eeq
which relates the OPE coefficients to an integral of the DIS amplitude.
By unitarity, we have  ${\rm Im}[A(x, q^2)]\geq 0$, so that the moments $a_s$ are positive and 
$ \tilde{c}_s^{(*)}  d_{s}^{(*)}\ge 0$. 
 Moreover, we must have $a_s > a_{s+2}$, which implies 
\beq\label{eq:convex}
\tau_s^{(*)} < \tau_{s+2}^{(*)}\,,
\eeq i.e.~the minimum twist is a non-decreasing function of spin. 

\subsection{Sum Rule in Holographic QFTs}
\label{sec:holo_sum}

So far, our discussion has been valid for general QFTs. We will now specialize to QFTs with holographic duals.  In particular, we consider a general scalar operator $\O$ of dimension $\Delta$ in a large-$N$ theory. The $N=\infty$ limit is a mean-field theory, where we can use conformal invariance, but the interactions may break conformal symmetry.  Following Komargodski et al.~\cite{Komargodski:2016gci}, we now construct the DIS amplitude for a such a QFT using the momentum state:
\beq
| P \rangle = \int \d^d y \, e^{i P \cdot y}\, \O(y) |0\rangle \equiv \O(P) |0 \rangle \,.
\eeq	
The  DIS amplitude (\ref{equ:DIS}) then reduces to a four-point function\hskip 1pt\footnote{These correlators include only the connected piece. The disconnected term has zero imaginary part.  Also, in this section and below, we perform the Fourier transform in Euclidean signature. Equation \eqref{equ:DIS} was in Lorentzian signature, so it differs by a factor of $i$.}
\beq\label{eq:fourpoint}
\begin{aligned}
A(q, P) 
&= \int \d^d y\, e^{-i q\cdot y}\, \langle 0 | \O(-P) \,{\mathbb T}(\O(y) \O(0) )\hskip 2pt \O(P) | 0 \rangle \\
&=  \langle \O(-P) \O(q) \O(-q) \O(P) \rangle' \, .
\end{aligned}
\eeq
In this section, we first show how to extract the spin-2 anomalous dimension, $\gamma_2$, from this four-point function and then translate the sum rule for the spin-2 moment of the DIS amplitude into a sum rule for $\gamma_2$.  In the next section, we will compute the four-point function holographically in terms of bulk interactions and use this to obtain a sum rule for certain parameters of the EFT of holographic RG flow. 

\vskip 4pt
First, we note that the operator $\O$ is normalized such that
\begin{align}\label{green}
\langle \O(x) \O(y) \rangle &= C_\O\,  |x-y|^{-2\Delta} \, ,\\[4pt]
\langle \O(p) \O(-p)\rangle' &= A_\O\, p^{2\Delta-d} \, ,  \qquad {\rm with} \quad A_\O = 2^{d-2\Delta}\pi^{d/2}   \frac{\Gamma(d/2-\Delta)}{\Gamma(\Delta)} \,C_\O\, . \label{equ:AO}
\end{align}
We are interested in the contribution of the following spin-2 composite operator
\beq\label{spin2def}
\O^{(2)}_{\alpha\beta} \equiv  P_{\alpha\beta}^{\gamma \delta}\big[\O \partial_{\gamma} \partial_{\delta} \O - b_\Delta \partial_\gamma \O \partial_{\delta} \O\big] \, ,
\eeq
where $P_{\alpha\beta}^{\gamma \delta} = \delta_\alpha^\gamma \delta_\beta^\delta - d^{-1}g_{\alpha\beta}g^{\gamma \delta}$ is the projector onto the traceless part. The relative coefficient in (\ref{spin2def}) is set by requiring $\O^{(2)}_{\alpha\beta}$ to be a conformal primary in the $N=\infty$ theory, which leads to\hskip 1pt\footnote{The simplest way to derive this is to compute $\langle \O^{(2)}_{\alpha\beta}(x) \O^2(y)\rangle$ by Wick contractions and requiring  it to vanish. Equivalently, \eqref{spin2def} is the specific linear combination that is annihilated by the special conformal generator $K_\alpha$.}
\beq
b_\Delta = \frac{\Delta+1}{\Delta} \, . \label{equ:bD}
\eeq
Assuming that the spin-2 operator has scaling dimension $\Delta_2$, it appears in the OPE as 
\beq\label{ooope}
\O(y)\O(0)  = \cdots + c_2^{(0)} |y|^{\Delta_2-2\Delta-2} y^\alpha y^\beta \O^{(2)}_{\alpha\beta}(0) + \cdots\,.
\eeq
At large $N$, composite operators have small anomalous dimensions, so $\Delta_2 = 2\Delta + 2  + \gamma_2$, with $\gamma_2 \ll 1$. The OPE coefficient $c_2^{(0)}$ is determined by comparison with the Taylor expansion,
\begin{align}
\O(y)  &= y \cdot \partial \O(0) + \frac{1}{2} (y\cdot \partial)^2 \O(0) + \cdots\\
\O(y)\O(0) &\supset \frac{1}{2} y^\alpha y^\beta \O \partial_\alpha \partial_\beta \O =c_2^{(0)} y^\alpha y^\beta \O^{(2)}_{\alpha\beta} + c' y^\alpha y^\beta \partial_\alpha \partial_\beta (\O^2) \, ,
\end{align}
where the last equality sets
\beq
c_2^{(0)}= [2(b_\Delta + 1)]^{-1} = \frac{\Delta}{2(2\Delta+1)}\,.
\eeq  
Finally, Wick contractions using the Green's function \eqref{green} lead to
\beq
\langle P | \O^{(2)}_{\alpha\beta}(0) |P\rangle = -\frac{1}{c_2^{(0)}} A_\O^2 \,P^{\gamma \delta}_{\alpha\beta} P_{\gamma}P_{\delta} P^{4\Delta-2d}\, .
\label{equ:2pt}
\eeq
The overall minus sign arises from the Fourier transform of the derivatives.  This sign is ultimately responsible for connecting the positivity of amplitudes with negativity of anomalous dimensions.  
The bulk interactions are responsible for a number of $1/N$ corrections, but at leading order in $1/N$, only the anomalous dimension $\gamma_2$ contributes to the DIS amplitude  
\beq
A(q,P) \supset - \gamma_2 \, c(d) A_\O^2\, P^{4\Delta-2d} \,   q^{-d-4} \left[ (P\cdot q)^2 - \frac{1}{d} P^2 q^2 \right]  , \label{equ:spin2}
\eeq
with $c(d) \equiv 2^{d-1}  d(d+2)\pi^{d/2}\,\Gamma({d/2})$. Note that $\gamma_2 \ll 1$ appears as a prefactor in the amplitude after taking the Fourier transform of~(\ref{ooope}).  One could have anticipated this from~(\ref{ooope}) as the OPE becomes analytic in $y$ when $\gamma_2 = 0$.  As a result, the Fourier transform must be a $\delta$-function of $q$ when $\gamma_2=0$ and the $q^{-d-4}$ contribution in~(\ref{equ:spin2}) must vanish as $\gamma_2 \to 0$.  This can be seen directly from expanding denominator of~(\ref{momope}) in $\gamma_2 \ll 1$.  In Appendix~\ref{app:betafunction}, we show that the appearance of $\gamma_2$ in~(\ref{equ:spin2}) is not a coincidence and follows from an explicit one-loop calculation of the anomalous dimension.

\vskip 4pt
Starting from the amplitude, one finds that the spin-2 moment is given by
\beq
a_2(q^2) =  -\frac{1}{4} \gamma_2\,  c(d) A_\O^2\, P^{4\Delta-2d}\,  q^{-d} \, .
\label{equ:a2}
\eeq 
This is the small-$\gamma$ expansion of the general result for the second moment in \eqref{equ:SUM0}. (There is no $\log q$ due to a cancellation with a gamma function from \eqref{momope}.) 
The result \eqref{equ:moment} then implies the following sum rule for the spin-2 anomalous dimension:
\beq\label{equ:SumRule}
\gamma_2 = -\frac{8}{\pi} \frac{1}{c(d) A_\O^2} \frac{1}{P^{4\Delta-2d}}  \lim_{q \to \infty} q^d \int_0^1 \d x \, x \,  \mbox{Im}[A(x,q^2)]\, ,
\eeq
where  the limit $q \to \infty$ has been applied to isolate the minimum-twist operator from the higher-twist operators. In practice, our effective description has a UV cutoff $\Lambda$ and the limit $q^2 \to \infty$ means $q^2 \to \Lambda^2$.

\vskip 4pt
Of special interest will be the case where $\O$ is proportional to the trace of the stress tensor in $d=3$ dimensions. Setting $\Delta=d=3$,  
we have $c(3)= 30 \pi^2$ and hence find
\begin{eBox}
\beq
\label{equ:SumRule2}
\gamma_2 = -\frac{4}{15\pi^3} \frac{1}{A_\O^2} \frac{1}{P^{6}} \lim_{q \to \infty} q^3 \int_0^1 \d x \, x \,  \mbox{Im}[A(x,q^2)]\, .
\eeq
\end{eBox}

\vskip 6pt 
\noindent
Because ${\rm Im}[A(x, q^2)] \ge 0$, must have $\gamma_2 \le 0$.
Moreover, in the limit $\gamma_{2} \to 0$, we must have $\gamma_{s>2} = 0$, because of the twist is a convex function of the spin, cf.~(\ref{eq:convex}). We will explore this further in \S\ref{sec:implications}.

\subsection{Matching Bulk and Boundary}
\label{equ:matching}

\begin{figure}[t!]
   \centering
      \includegraphics[scale=1.]{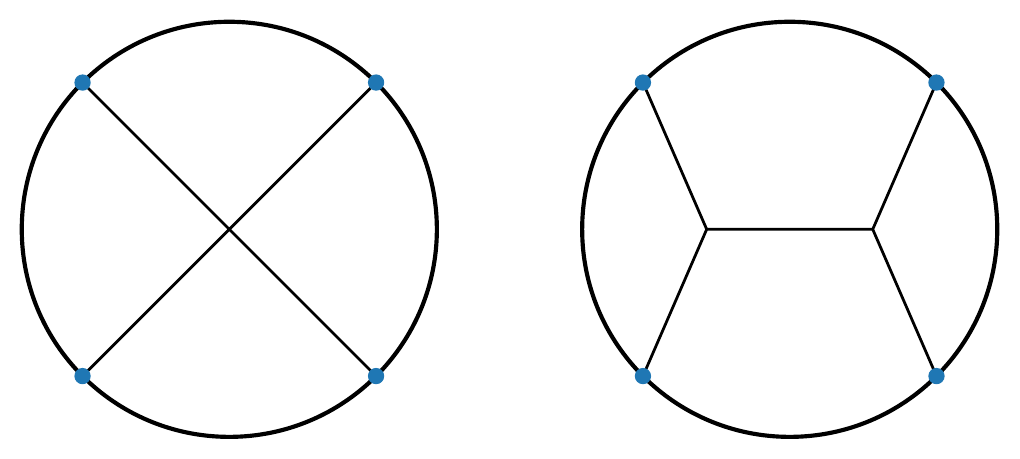} 
      \caption{Witten diagrams associated with the contact interaction $( \partial \pi_c)^4$ and the exchange interactions arising from $\partial_r \pi_c(\partial \pi_c)^2$ and $(\partial_r \pi_c)^3$.}
      \label{fig:Witten}
\end{figure}

We will now translate (\ref{equ:SumRule2}) into a constraint on the EFT of holographic RG flow.  
The upshot of the previous section is that, for $d=3$, we are looking for all contributions to the four-point function that behave as $(P \cdot q)^2 P^6 /q^7$. At tree level, there are two types of Witten diagrams to consider, contact and exchange diagrams (see Fig.~\ref{fig:Witten}).  The contact diagrams are the easiest to understand, so we will consider these first.

\vskip 4pt
For any contact interactions of the form shown in Fig.~\ref{fig:Witten}, all propagators carry a momentum $P$ or $q$ and thus do not depend on $P\cdot q$.  Any nontrivial $x$-dependence must therefore arise from the interaction vertex itself, namely from derivatives acting on the external lines.  It is easy to see that producing a spin-$s$ moment of the DIS amplitude requires a vertex with at least $2s$ transverse derivatives.  
In any dimension, only a single contact interaction in the Lagrangian~(\ref{equ:Lint3}) contributes to the spin-2 DIS amplitude:
\beq
{\cal L}_{\rm int} \,\subset\, \frac{1}{4!} \lambda_4\, \frac{1}{f_\pi^{d+1}} \, a^{-4} ( \partial \pi_c)^4\,, \label{equ:p4}
\eeq
where $\lambda_4 \equiv 3 (1-c_r^2)$.   We are interested in the DIS amplitude for the trace of the stress tensor.  Holographically, the operator dual to $T$ is the scalar metric perturbation $\zeta$, which is related to the canonically-normalized Goldstone mode by $\pi_c \approx - f_\pi^{(d+1)/2}\, \zeta \, /H$.
The full calculation is presented in Appendix~\ref{app:holographic}.  For $d=3$, the result is
\beq
A(q, P) \supset \frac{55}{24}  \lambda_4\, \frac{f_\pi^{4}}{H^{4}}\, \frac{ (2 P\cdot q)^2 P^6}{q^7} \, .
\label{equ:contact}
\eeq
This contribution is not manifestly positive, but gives the same answer (up to factors of $f_\pi /H$) as the covariant $(\nabla \phi)^4$ interaction in pure AdS.  The reason for the agreement is that the additional contact terms containing $r$-derivatives, which are needed to make the interaction covariant, do not contribute to the spin-2 moment. Demanding that this contribution alone is positive forces $\lambda_4 \geq 0$~\cite{Hartman:2015lfa}, in agreement with the flat-space result~\cite{Adams:2006sv}.

\vskip 6pt
The contributions to the DIS amplitude from exchange diagrams with cubic vertices are significantly more involved than  the contact interaction.  The reason is that the internal bulk-bulk propagator carries a momentum $q\pm P$, which leads to an infinite series in $x^{-s}$ when expanded in $P \ll q$.  As a result, the contribution to the spin-2 moment is not isolated to a particular combination of contractions of derivatives on the external lines.

\vskip 4pt
It is well known that, for large twist, the spin-$s$ anomalous dimensions computed in AdS are directly related to the corresponding spin-$s$ partial waves of the flat-space scattering amplitude in the bulk (e.g.~\cite{Fitzpatrick:2010zm}).  However, at finite twist, such a connection is not guaranteed and indeed the exchange diagrams that contribute to $\gamma_2$ do not appear in the spin-2 partial wave 
of the flat-space amplitude computed in~\cite{Baumann:2015nta}.  The origin of this difference is that, in flat space, the spin-2 component of the amplitude from exchange diagrams cancels in the sum over contractions. These cancelations depend crucially on the flat-space propagator and do not arise for the AdS exchange diagrams.  As a result, we find that the spin-2 contribution to the DIS amplitude in the dual QFT is related to a spin-0 amplitude of the flat-space S-matrix.  

\vskip 4pt
At leading order in derivatives, the EFT of slow RG flow contains only
 two cubic interactions
\beq
{\cal L}_{\rm int} = \frac{1}{3!} \frac{1}{f_\pi^{(d+1)/2}} \left[ \lambda_3 \, a^{-2}  \partial_{\tilde r} \pi_c \, (\partial \pi_c)^2 +  \tilde \lambda_3\, (\partial_{\tilde r} \pi_c)^3 \right] ,
\eeq
where $\lambda_3 \equiv 12 c_2/c_r$ and $ \tilde \lambda_3 \equiv (12 c_2 + 8 c_3)/ c_r^3$.  This allows for three distinct exchange contributions to the DIS amplitude, which are computed explicitly in Appendix~\ref{app:holographic}. 
For $d=3$, the result is
\beq
A(q, P) \supset \left[\frac{495}{512} \tilde{\lambda}_3^2 - \frac{605}{256} \lambda_3\tilde \lambda_3 + \frac{6655}{4608}\lambda_3^2\right] \frac{f_\pi^4}{H^4}   \frac{(2P\cdot q)^2 P^6}{q^7}\,. \label{equ:exchange}
\eeq
The signs in this expression are nontrivial and important, given that $A(q,P)$ is required to be positive by unitarity.  Furthermore, unlike for the contact contribution, the sign of the first and third terms cannot be absorbed into the signs of the couplings. 

\vskip 6pt
Adding (\ref{equ:contact}) and (\ref{equ:exchange}), we find that the total DIS amplitude is
\beq
A(q, P) \supset  \left[\frac{55}{24}\lambda_4 + \frac{495}{512} \tilde{\lambda}_3^2 - \frac{605}{256} \lambda_3\tilde \lambda_3 + \frac{6655}{4608}\lambda_3^2\right] \frac{f_\pi^{4}}{H^{4}} \frac{(2P\cdot q)^2 P^6}{q^7} \, .
\eeq
Recalling from \S\ref{sec:dual} that $T$ is normalized in $d=3$ such that $A_{{\cal O} = T}  = {c_r^2 f_\pi^4/\tilde H^4}$~\cite{Maldacena:2002vr}, we use~(\ref{equ:spin2}) to obtain
\beq
\begin{aligned}
\gamma_2 &= - \frac{1}{\pi^2}\left(\frac{\tilde H}{f_\pi}\right)^{4} \left(  \frac{11}{36}\lambda_4 + \frac{33}{256} \tilde \lambda_3^2 - \frac{121}{384} \lambda_3 \tilde \lambda_3 + \frac{1331}{6912} \lambda_3^2 \right) \\
& =  - \frac{1}{\pi^2} \left(\frac{\tilde H}{f_\pi}\right)^{4} \left[ \, \frac{11}{12}  (1-c_r^2)  + \frac{33}{4 c_r^6} \left(\frac{3}{2} c_2 + c_3\right)^2  - \frac{121}{4 c_r^4} c_2 \left(\frac{3}{2} c_2 + c_3\right) + \frac{1331}{48 c_r^2} c_2^2   \right] , \label{eq:final_amp}
\end{aligned}
\eeq
where,  in the second equality, we used the relations between $\lambda_3$, $\tilde \lambda_3$, $\lambda_4$ and $c_r$, $c_2$, $c_3$.    
Using (\ref{equ:SumRule}), we arrive at the following sum rule 
\begin{eBox}
\beq
\label{eq:sumrulefinal}
\left(1-c_r^2\right) + f(c_2,c_3) \,=\,
\frac{1}{55 \pi} \left(\frac{f_\pi}{\tilde H}\right)^{4} \lim_{q \to \infty} q^3 \int_0^1 \d x\, x \  {\rm Im}[A(x, q^2)] \, ,
\eeq
\end{eBox}
where $\tilde H =c_r H$ is the effective curvature scale and we have defined
\begin{align}
f(c_2,c_3) 
&\equiv \frac{1}{c_r^6} \Bigg( 3 \left(\frac{3}{2} c_2 + c_3\right) - \frac{11}{2} c_r^2 c_2 \Bigg)^2 \geq 0\, .
\end{align}
The fact that this expression organizes into a single perfect square is nontrivial. 
We note that the left-hand side of \eqref{eq:sumrulefinal} is manifestly positive for $0 < c_r \leq 1$. We therefore do not find any surprising constraints on the couplings of the EFT from positivity of the DIS amplitude. The power of the sum rule is twofold: First, given the IR quantities on the left, we can infer properties of the UV amplitude on the right.  Second, the sum rule can be recycled to constrain an infinite tower of higher-derivative operators through~(\ref{equ:moment}).

\subsection{Freedom at the Speed of Light}
\label{sec:implications}

It is natural to expect our sum rule to be particularly illuminating when the propagation speed approaches the speed of light, $c_r \to 1$.  Bulk causality forbids $c_r >1$ and sum rules usually simplify when positivity constraints are saturated.  In the context of inflation, this expectation motivated the conjecture in~\cite{Baumann:2015nta} that $c_s=1$ should be a free theory (in the decoupling limit). We will now establish a similar result for the holographic RG flow.

\vskip 4pt
Evaluating the sum rule~(\ref{eq:sumrulefinal}) for $c_r =1$, we get 
\beq
9\hskip 1pt c_3 ^2 = \frac{1}{55 \pi} \left(\frac{f_\pi}{\tilde H}\right)^{4} \lim_{q \to \infty} q^3 \int_0^1 \d x\, x \,  {\rm Im}[A(x, q^2)] \, . \label{equ:SR}
\eeq
This constraint alone is still consistent with a wide range of  interacting EFTs.  However, bulk causality requires $c_3 = 0$ when $c_r =1$. 
To see this, we change the background by writing $\pi = \alpha r + \tilde \pi$, for some $\alpha \ll 1$.  The infall speed for $\tilde \pi$ then is $\tilde c_r = 1-8 \alpha \hskip 1pt c_3$.  Since $\alpha$ can have any sign this implies superluminal propagation\footnote{In a relativistic theory, the commutator of two operators must vanish outside the light cone in any state of the theory.  We can therefore test the consistency of the theory by expanding around nontrivial backgrounds~\cite{Adams:2006sv}, where the propagation speed is a direct reflection of the commutator and hence of causality. } unless $c_3=0$.  Expanding the action to order $\alpha^2$, one would furthermore find that $c_4 \geq 0$.  
Imposing the additional constraint $c_3=0$ on (\ref{equ:SR}) forces the integrated DIS amplitude to vanish at high energies,
\beq\label{eq:cr1sum}
 \lim_{q \to \infty} q^3 \, \int_0^1 \d x\, x\,  {\rm Im}[A(x, q^2)]  = 0 \, .
\eeq
This constraint on the UV behavior feeds back into the EFT through the higher-spin DIS amplitudes.  Up to a power of $x$, the integrand in~(\ref{eq:cr1sum}) controls all of the moments $a_s(q^2)$, for $s\geq 2$.  Specifically, using~(\ref{eq:cr1sum}) and positivity of the cross section ${\rm Im}[A(x, q^2)] \ge 0$, for $c_r =1$, we have 
\beq\label{eq:ascr1}
\lim_{q \to \infty}  q^3 a_s(q^2)  =\  \lim_{q \to 0} q^3 \frac{2}{\pi}\int_0^1 \d x\, x^{s-1}\, {\rm Im}[A(x, q^2)] = 0\, .
\eeq
These higher-spin DIS amplitudes get non-zero contributions in the bulk EFT from contact interactions of the form
\beq\label{eq:higher_spin}
{\cal L}_{\rm int}  \supset \partial_{\alpha_1 \ldots \alpha_{i}} \pi_c \hskip 2pt \partial^{\alpha_1 \ldots \alpha_{i}} \pi_c \hskip 2pt  \partial_{\alpha_{i+1} \ldots \alpha_{s}} \pi_c \hskip 2pt  \partial^{\alpha_{i+1} \ldots \alpha_{s}} \pi_c \, .
\eeq
By construction, the contractions in the Witten diagram give either $(P\cdot q)^s$ or $(P\cdot q)^0$ and therefore these interactions do not contribute to $a_2(q^2)$. This means that all such couplings must vanish if $c_r=1$.  

\vskip 4pt
The argument for the vanishing of higher-derivative operators can be extended from $c_r=1$ to the limit $c_r \to1$.  Causality constraints around nontrivial backgrounds can be applied when $0< (1-c_r^2) \ll 1$ if we impose that the gradient satisfies $\alpha \lesssim 1$, so that we are within the regime of validity of the EFT.  Forbidding superluminal propagation in the radial direction then requires $|c_3| \lesssim (1-c_r^2)$, so the sum rule becomes
\beq\label{eq:smallcr}
\left(1-c_r^2\right) + {\cal O}\left((1-c_r^2)^2\right) \,=\,
\frac{1}{55 \pi} \left(\frac{f_\pi}{\tilde H}\right)^{4} \lim_{q \to \infty} q^3 \int_0^1 \d x\, x \,  {\rm Im}[A(x, q^2)] \, ,
\eeq
as $c_r \to 1$.  Furthermore, combining this constraint with~(\ref{eq:ascr1}) bounds the strength of an infinite tower of couplings in terms of $1-c_r^2$.

\vskip 4pt
The above results provide support for the conjecture that $c_r=1$ is a free theory (in the decoupling limit).  The sum rule and bulk causality force both an infinite number of operators in the EFT to vanish and  the DIS amplitude to become trivial in the high-energy limit.  However, not all operators of the EFT are constrained in this way.  In particular, the operators associated with $M_{n>3}$ do not contribute to $a_{s\geq 2}(q^2)$ or ${\rm Im}[A(P,q)]$ at this order in $ H/f_\pi$ and therefore are do not appear on either side of the sum rule.  However, the presence of such operators would still produce nontrivial higher-point amplitudes in the QFT and thus would be indirectly constrained by the restrictions on the DIS amplitude.  A similar situation arises in energy correlators for certain extremal limits. It was shown in~\cite{Zhiboedov:2013opa} that such theories are free by bounding the size of $(n+1)$-point energy correlators in terms of $n$-point correlators.  

\subsection{Implications for Slow Motion}
\label{sec:smallcr}

It is also interesting to explore the meaning of the sum rule in the limit $c_r \ll 1$, where the interactions become large.  For generic $c_3$, the sum rule in (\ref{eq:sumrulefinal}) then takes the form
\beq
f(c_2,c_3) = \frac{1}{c_r^6}   \left(\frac{3}{2} c_2 + c_3\right)^2 + {\cal O}(c_r^{-2}) \,=\, \frac{1}{495 \pi} \left(\frac{f_\pi}{\tilde H}\right)^{4} \lim_{q \to \infty} q^3 \int_0^1 \d x\, x \  {\rm Im}[A(x, q^2)] \, .
\label{equ:SUM}
\eeq
Self-consistency of this sum rule imposes a constraint on the DIS amplitude, including its high-energy behavior.  In particular, the limit on the right-hand side of (\ref{equ:SUM}) should be finite, so the integrated amplitude must scale as $q^{-3}$ for $q \to \infty$.   The precise meaning of this constraint is somewhat unclear.  
 Curiously, the best known UV-complete model of small $c_r$, the DBI action, does not match the $c_r^{-6}$ scaling displayed in (\ref{equ:SUM}).    
 In this case, the parameters $c_3$ and $c_2$ are related as $\frac{3}{2} c_2 + c_3 = \frac{3}{8} c_r^2(1-c_r^2)$ (see Appendix~\ref{app:DBI}), so that 
\begin{align}
f(c_2,c_3) = \frac{1}{16} \frac{(1-c_r^2)^2}{c_r^{2}} \,  \qquad ({\rm DBI})\, ,
\end{align}
which scales as $c_r^{-2}$ rather than $c_r^{-6}$, for small $c_r$.  This implies that the UV behavior of the DIS amplitude dual to a bulk DBI theory is weaker at high energies than for a generic higher-derivative theory.  
Presumably the origin of this cancelation has a deeper origin in the form of the UV-completeness of DBI.  

  \vskip 4pt
Finally, at small $c_r$, the convexity of the minimum-twist anomalous dimensions bounds the higher-spin anomalous dimensions $\gamma_{s>2}$ in terms of $\gamma_2$.  These higher-spin anomalous dimensions can be related to couplings in the bulk EFT in the same way as in the case of spin~2, the only difference being the number of $q^\alpha$-derivatives that must appear for the contact diagrams. Exchange diagrams contribute to the entire tower of minimal-twist anomalous dimensions but, at least for $c_2$ and $c_3$, are consistent with $\gamma_{s} < \gamma_{s+2}$ in the absence of additional contact interactions.  For small $c_r$, these higher-derivative operators may also  
modify the dispersion relation of the fluctuations with higher powers of momenta, $\omega =c_s k + k^2/\rho +\cdots$. Convexity in spin ensures that higher-derivative interactions cannot be parametrically larger than the interactions associated with $c_2$ and $c_3$.  A similar statement for the parameters of the EFT of inflation would have important observational implications.

\section{From RG Flow to Inflation}
\label{sec:dS}

\begin{figure}[h!]
   \centering
      \includegraphics[scale=1.]{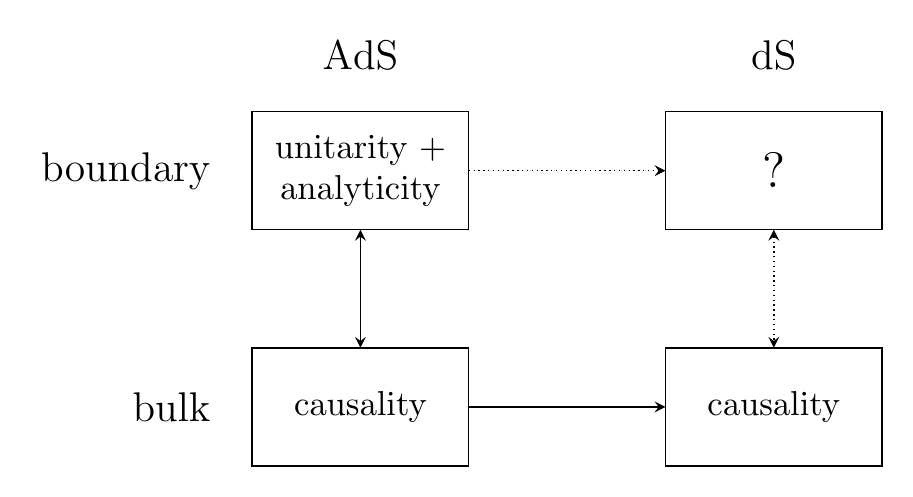} 
      \caption{Connections between bulk causality and boundary unitarity in AdS and dS spacetimes.  On the AdS side, unitarity and analyticity arguments can be formulated rigorously on the boundary and map to solid causality constraints in the bulk. On the dS side, on the other hand, the rules of the boundary QFT are much less well established.
However, existing causality constraints in AdS can be mapped to equivalent constraints in dS.  We therefore anticipate that boundary constraints derived in AdS will apply in dS without necessarily defining an equivalent boundary description.   }
      \label{fig:logic}
\end{figure}

The sum rule derived in the previous section relied heavily on the dual QFT being relativistic and unitary. The absence of a similar dual field theory in de Sitter space might seem to imply that the constraints derived here have no counterpart in cosmology.  On the other hand, holography demands that our sum rule has a bulk interpretation.  Indeed, when the bulk is described by pure AdS, similar constraints on the dual CFT~\cite{Hartman:2015lfa, Afkhami-Jeddi:2016ntf,Cordova:2017zej} are related to sub-horizon physics and can be derived from bulk causality~\cite{Adams:2006sv,Camanho:2014apa}. 
Since local causality arguments are relatively insensitive to the asymptotic structure of the spacetime, they translate more readily between AdS and dS.
Hence, if our sum rule can be related to local causality constraints in the bulk, it may be easier to see what it could imply for inflation.   This path from AdS to dS, 
illustrated in Fig.~\ref{fig:logic}, is similar to the logic that gives rise to the ``quasi-bounds" on graviton couplings obtained in~\cite{Camanho:2014apa, Cordova:2017zej} and has also appeared in the connection between AdS and holographic cosmology in~\cite{McFadden:2009fg}.  

\vskip 4pt
In this section, we will first show in what sense our sum rule is (and is not) a consequence of bulk causality.  In particular, we will explain that the sign constraints implied by the sum rule are related to superluminal propagation in the transverse directions, but do not capture causality in the radial direction.  We will then show that the bulk causality constraints found in AdS for both the transverse and radial directions apply equally to the equivalent couplings in the EFT of inflation.  

\subsection{Relation to Bulk Causality in AdS}

It is natural to think that the sign constraint implied by our sum rule has an obvious interpretation from causality, namely $c_r \leq 1$.  Certainly, if we impose causality in this way, the amplitude is always positive, as required by unitarity.  On the other hand, the sum rule alone does not imply this constraint on $c_r$, nor does it capture causality constraints on other parameters.  As explained in \S\ref{sec:implications}, forbidding superluminal propagation around nontrivial backgrounds requires $c_3=0$ and  $c_4 \geq 0$,  when $c_r =1$. Yet, in the DIS amplitude, the contribution to from $c_3$ is always positive and $c_4$ does not appear at all.  

\vskip 4pt
A likely explanation for this mismatch between the bulk causality constraints and the boundary DIS amplitude is that the missing bulk constraints arise from superluminal propagation in the \textit{radial} direction, while the DIS amplitude is only sensitive to propagation in the \textit{transverse} directions.  A similar phenomenon has been observed in the context of the conformal collider bounds on stress-tensor OPE coefficients in CFTs~\cite{Hofman:2008ar}. In the bulk, the collider bounds are derived from causality in the transverse directions \cite{Hofman:2009ug}, while, on the boundary, they can be derived either by DIS~\cite{Komargodski:2016gci} or by microcausality of local operators in the CFT \cite{Hartman:2016dxc,Hofman:2016awc}. On the other hand, stronger causality bounds were derived in \cite{Camanho:2014apa} from propagation in the radial direction. These bounds can also be derived using CFT techniques~\cite{Afkhami-Jeddi:2016ntf,Kulaxizi:2017ixa}, but only with more elaborate sum rules that take advantage of the special structure of holographic CFTs, including large $N$ and a large gap in the spectrum of higher-spin operators.  

\vskip 4pt
There is a sharp connection between the positivity of our sum rule and propagation speeds in the transverse directions, even though the sum rule constrains the radial speed~$c_r$.   Recall that when $c_r \approx 1$, our sum rule simplified to~(\ref{eq:smallcr}), so that the positivity of the amplitude implies the constraint $c_r \leq 1$.  
This can also be derived from transverse causality alone. Recall that, for $c_r \approx 1$, only $\lambda_4 (\partial \pi)^4$ contributes to the sum rule.  Expanding $\pi$ around a gradient in any transverse direction, $\pi = \alpha_\beta x^\beta + \tilde \pi$, one then finds superluminal propagation unless $\lambda_4 \geq 0$~\cite{Adams:2006sv}.  Since $\lambda_4 \propto 1-c_r^2$, this implies $c_r \leq 1$. Although this is a bound on the radial propagation speed, we see that it fundamentally came from causality in the transverse directions.

\subsection{From Bulk Causality to Inflation}
\label{sec:bulk_dS}

We have seen that bulk causality in the EFT of RG flow plays a crucial role in the structure of the sum rule.  In particular, the result that the theory becomes free when the propagation speed approaches the speed of light is only implied when the sum rule is combined with bulk causality constraints in the radial direction.  The EFT of inflation is subject to precisely the same causality constraints derived from expanding around nontrivial backgrounds~\cite{Baumann:2015nta}.  In particular, in the limit $c_s \to 1$, the Lagrangian takes the form
\begin{align}
\tilde{\cal L}=  -\frac{1}{2}( \partial \pi_c)^2 -\frac{2}{f_\pi^{2}} \left[  \frac{2}{3}c_3 \dot \pi_c^3 \right] + \frac{1}{2 f_\pi^{4}} \left[4 c_3 \dot \pi_c^2 ( \partial \pi_c)^2 + \frac{4}{3}c_4 \dot \pi_c^4\right]  .
\end{align}
Expanding around $\pi_c =-\alpha f_\pi^2 t + \tilde \pi_c$, the propagation speed for $\tilde \pi_c$ becomes $c_s^2 = 1+ 8 \alpha \hskip 1pt c_3$ (to linear order in $\alpha$).  Since $\alpha$ can have either sign, subluminality then requires $c_3 =0$.  Setting $c_3= 0$ and expanding to order $\alpha^2$ gives $c_s^2 = 1-4 \alpha^2 c_4$, which implies $c_4\geq 0$.  Furthermore, if we could add an interaction of the form  $ \lambda_4 (\partial_i \pi)^4$, while keeping $c_s=1$, expanding around a gradient in the transverse direction, $\pi = \alpha_i x^i + \tilde \pi$, would again require $\lambda_4 \geq 0$.  The equivalence between the causality constraints in these two EFTs provides circumstantial evidence that freedom at the speed of light is also a property of the EFT of inflation, as originally conjectured in~\cite{Baumann:2015nta}. 
 
\vskip 4pt
A critical difference between holographic RG and inflation is that for inflation there is no unitarity QFT dual in which to define the sum rule.  Nevertheless, all quantities appearing in the AdS sum rule were calculated from a bulk four-point function.  We can calculate an equivalent inflationary four-point function at the future boundary of dS using ${\cal A} \equiv \delta^4 \Psi/\delta \zeta^4$, where $\Psi$ is the wavefunction of the universe \cite{Maldacena:2002vr}.  Although this is not a typical (in-in) cosmological correlator, it is still a quantity we can calculate using the effective theory of inflation.  Following the same procedure as in AdS, we are then led to the following sum rule 
\beq
(c_s^{-2} -1) + f(c_2,c_3) \,\propto\, \lim_{q\to \infty} q^3 \oint \d x\,  x\,  {\cal A}(x,q^2) \, ,
\eeq
where $(1-c_r^2) \to (c_s^{-2} -1)$ is just the map from $c_r \to c_s$ and $f(c_2, c_3)$ is the contribution from exchange diagrams in de Sitter. In principle, this relates observable quantities to the UV, but unfortunately, 
the analytic properties of the de Sitter four-point function, ${\cal A}(x,q^2)$, are not known, so that positivity is not guaranteed.

\vskip 4pt
A more direct approach, taken in~\cite{Baumann:2015nta}, is to use bulk scattering to derive sum rules for the EFT parameters.  This approach can be applied to either EFT, but in both cases positivity of the sum rule has not been established without Lorentz invariance.  Nevertheless, using the two-to-two amplitude of the Goldstone boson in the forward limit, ${\cal A}(s)$, a sum rule can be derived for the couplings of the EFT of inflation~\cite{Baumann:2015nta} 
\beq
(c_4+1) -\big[ (2 c_3+1)-a(c_s)\big]^2 - b(c_s) = \frac{\Lambda^4}{\pi} \int_{-\infty}^{\infty} \d s\, \frac{{\rm Im}[{\cal A}(s)]}{s^3}\, , \label{equ:B2015}
\eeq
where $a(s)$ and $b(c_s)$ are positive functions of $c_s<1$ and vanish at $c_s=1$.  Unlike in relativistic theories, the right-hand side of (\ref{equ:B2015}) is not known to be positive, mainly because there is no non-relativistic analogue of crossing symmetry to constrain the integral for negative $s$.  Despite this caveat, in specific examples the right-hand side is positive, in which case this expression puts constraints the amplitude and sign of $c_4$, in terms of $c_3$ and $c_s$.
 
 \newpage
\section{Conclusions}
\label{sec:conclusions}

Understanding the space of inflationary models and renormalization group flows are two central problems in fundamental physics.  In both cases, the vast space of theories presents a serious challenge for organizing and classifying the possibilities.  
 Inflation describes a time-dependent gravitational background, where dynamical effects such as particle production, thermalization and even eternal inflation may occur as part of the inflationary history.  In contrast, RG flows can be nontrivial in the vacuum state and describing the range of possibilities remains an open problem even for relativistic theories in flat space. 
Nevertheless, we expect the space of viable inflationary models and RG flows to satisfy a common set of basic principles such as locality, causality and unitarity.   In some cases, these principles can be distilled into sum rules which directly relate the long-distance phenomenology to the ultimate microscopic building blocks.  Revealing new and unexpected connections in the properties of viable theories is a common goal for both fields.   

\vskip 4pt
Holography provides an interesting connection between inflation and special types of slow RG flow. Concretely, the EFT of Goldstone fluctuations during inflation~\cite{Cheung:2007st} is in one-to-one correspondence with a similar EFT in AdS~\cite{Kaplan:2014dia}, where it maps to an RG flow in a relativistic and unitary QFT.
In this paper, we explored the relationship between these effective field theories in more detail. 
We obtained new constraints on the parameters that controls the speed of propagation of scalar metric perturbations, called $c_s$ and $c_r$ in the two cases, respectively. 
The propagation speed $c_r$ has a natural interpretation on the boundary as the speed at which an operator, initially localized in space, spreads under time evolution.   
Using deep inelastic scattering in the dual QFT, we derived a sum rule for $c_r$ and showed that unitarity implies an interesting positivity constraint.
In the limit $c_r \to 1$, the sum rule requires an infinite number of higher-derivative operators to vanish, which suggests that the bulk theory becomes free.  
While an analogous sum rule for $c_s$ has been obtained for inflation~\cite{Baumann:2015nta}, it has  (so far)  not been possible to derive the same positivity properties.  
Such a positivity constraint would provide a valuable tool for interpreting the measurements of primordial non-Gaussianity from inflation, since it would turn any constraint on $c_s$ into a bound on an infinite number of other operators.

\vskip 4pt
More generally, the relation between the two EFTs studied in this paper suggests that a number of ideas developed in the context of inflation will have interesting analogues in RG flow.   A deeper understanding of the allowed space of RG flows could therefore significantly impact our view on viable inflationary models, and vice versa. For example, we observed that fairly conventional inflationary models correspond in AdS to nearly scale-but-not-conformal field theories.  Similarly, models like axion monodromy inflation~\cite{Flauger:2009ab,Slosar:2019gvt} would further motivate more exotic RG flows~\cite{Kiritsis:2016kog}, including discrete scale invariance~\cite{Behbahani:2011it,Flory:2017mal}, disorder~\cite{Green:2014xqa} and Anderson localization~\cite{Amin:2015ftc,Garcia:2019icv} along the RG flow.   Our hope is that the flow of information works in both directions, such that our knowledge of relativistic QFT can be used to derive insights into inflation and our knowledge of the space of inflationary models can motivate new questions about the space of interesting RG flows.

\newpage
\paragraph{Acknowledgements}
We are grateful to Ken Intriligator, Aneesh Manohar, Liam McAllister, John McGreevy, Rafael Porto, David Stefanyszyn, Amir Tajdini, and Matias Zaldarriaga for helpful conversations. 
We also thank the participants of the Simons Symposium ``Amplitudes Meet Cosmology" for insightful discussions and the Simons Foundation for funding the symposium. D.\,B.~and D.\,G.~thank the Gordon Research Conference ``New Physics in the Era of Precision Cosmology" for hospitality while this work was being completed.
D.\,B.~is supported by a Vidi grant of the Netherlands Organisation for Scientific Research~(NWO) that is funded by the Dutch Ministry of Education, Culture and Science~(OCW).  
D.\,G.~is supported by the US~Department of Energy under grant no.~DE-SC0019035.
T.\,H.~is supported by the US Department of Energy grant DE-SC0014123.

\clearpage
\appendix

\section{Anomalous Dimensions from Holography}
\label{app:holographic}

In this appendix, we compute the spin-2 moment of the DIS amplitude for the Goldstone action~(\ref{equ:Lint3}). 
The calculations will be performed in Poincar\'e coordinates, where the AdS metric takes the form
\beq
\d s^2 =\frac{\d z^2 + \ \d x_\alpha \d x^\alpha }{(\tilde H z)^2} \, . 
\eeq
The appearance of $\tilde H = c_r H$, rather than $H$, is a reminder that this is not the physical metric, but only the effective metric for the $\pi$-fluctuations in the $\tilde r$-coordinates introduced in \S\ref{sec:EFTofRG}.

\subsection{Contact Diagrams}
\label{app:contact}

Whether or not a contact interaction contributes to a spin-$s$ moment depends on the types of derivatives acting on the bulk-to-boundary propagators.  It is easy to see that only a single interaction in (\ref{equ:Lint3}), namely  $\tfrac{1}{4!} \lambda_4 (\partial \pi_c)^4$, contributes to the spin-2 moment. For this interaction, the DIS amplitude becomes
\beq
A(q, P) =  \frac{16}{4!} \lambda_4 \, c_r^4\left(\frac{ f_\pi}{\tilde H}\right)^{d+1}  \int \frac{\d z}{z^{d+1}} (P\cdot q)^2 z^4 \frac{ z^{2 d} K_\nu(q z)^2 K_\nu (P z)^2}{z_0^{2d} K_\nu(q z_0)^2 K_\nu (P z_0)^2} \, ,
\label{equ:DIS3}
\eeq
where $\nu \equiv \Delta - d/2$, the functions $K_\nu$ are Bessel functions arising from the bulk-to-boundary propagators, and $z_0$ is the regulator of the bulk AdS spacetime.  
Consistency of the amplitude requires that $q z_0 < 1$, but allows for $q z > 1$.  Since the integral in (\ref{equ:DIS3}) is exponentially suppressed for $q z >1$, it has most of its support in the region $P z \lesssim P / q \ll 1$.  We can therefore expand the integrand in powers of $P z \ll 1$, such that the amplitude becomes 
\beq
\begin{aligned}
A(q, P)
\approx & \frac{2}{3} \lambda_4 \,c_r^4\left(\frac{ f_\pi}{\tilde H}\right)^{d+1}\int \d z  \,(P\cdot q)^2 z^{d +3} K_\nu(q z)^2 \frac{1}{ z_0^{2d} (a_\nu (q z_0)^{-\nu} + b_\nu (q z_0)^{\nu} )^2 } \frac{z_0^{2 \nu}}{z^{2 \nu}} \\[2pt]
&\hspace{2cm} \qquad \times \left[ 1 +\cdots+ \frac{b_\nu}{a_\nu} P^{4 \nu} (z^{4\nu} - 4 z^{2\nu}z_0^{2\nu}+3 z_0^6)  \right] , \label{equ:AqP}
\end{aligned}
\eeq
where $K_\nu(x \to 0) = a_\nu x^{-\nu} [1+ {\cal O}(x^2)] + b_\nu x^{\nu}[1+ {\cal O}(x^2)]$. 
To extract the spin-2 component, we isolate the piece scaling as $P^{4\nu} q^{-d-4}$, cf.~(\ref{equ:spin2}).  
The relevant term in (\ref{equ:AqP}) is 
\begin{align}
A(q, P) &\supset \frac{2}{3} \lambda_4 \,c_r^4\left(\frac{ f_\pi}{\tilde H}\right)^{d+1} \, (P\cdot q)^2 P^{4\nu}  \int \d z  \,z^{d +3} K_\nu(q z)^2 \frac{(q z_0)^{2 \nu}}{z_0^{2 d} a_\nu^2} \frac{b_\nu^2}{a_\nu^2} z_0^{2\nu} z^{2\nu}  \, , \nonumber \\[2pt]
&= \frac{55}{24}  \lambda_4\, \frac{f_\pi^{4}}{H^{4}}\, \frac{ (2 P\cdot q)^2 P^6}{q^7} \, ,
\end{align}
where, in the second line, we have evaluated the integral for $d=3$.  It is important that we calculated the coefficient of $P^{4 \nu}$ in the limit $\nu \to 3/2$, to keep it distinct from semi-local terms that do not contribute to the anomalous dimension and scale as $P^6$ for general $d$ and $\nu$.  

\subsection{Exchange Diagrams}
\label{app:exchange}

Exchange diagrams involving two cubic vertices introduce an infinite series of terms containing powers of $P\cdot q$ arising from the momentum propagating in the internal line, namely $P\pm q$.  
This gives contributions to the DIS amplitude at all spins.  The spin-2 component is again extracted by isolating the terms scaling as $P^{4\nu} q^{-d-4}$.

\vskip 4pt
The Goldstone Lagrangian (\ref{equ:Lint3}) has two cubic interactions  
\beq
{\cal L}_{\rm int} = \frac{1}{3!} \frac{1}{f_\pi^{(d+1)/2}} \left[ \lambda_3 \, \partial_{\tilde r} \pi_c (\partial \pi_c)^2 +\tilde \lambda_3 \, (\partial_{\tilde r} \pi_c)^3 \right] ,
\label{equ:cubic}
\eeq
where $\lambda_3 \equiv 12 c_2/c_r$ and $\tilde \lambda_3 \equiv (12 c_2 + 8 c_3)/ c_r^3$.  
To compute the exchange contributions associated with these interactions, we need the following bulk-to-bulk  and  bulk-to-boundary propagators
\begin{align}
G_Q(z_1,z_2) &= \I_Q(z_1) \K_Q(z_2)\, , \quad z_2 > z_1\, ,\\[6pt]
F_Q(z) &= \frac{z^{d/2} K_\nu(Q z)}{z_0^{d/2} K_\nu(Q z_0)}\, ,
\end{align}
where $ \I_Q(z) = z^{d/2} I_\nu(Qz)$ and  $ \K_Q(z) = z^{d/2} K_\nu(Qz)$, with $Q$ labelling the norm of the momentum in the transverse directions. 

\vskip 4pt
The contribution proportional to $\tilde \lambda_3^2$ is the most straightforward to calculate since all contractions are equivalent:  
\beq
\begin{aligned}A(q, P)_{\tilde \lambda_3^2} &\,=\, 2 \tilde \lambda_3^2 \,c_r^4\left(\frac{ f_\pi}{\tilde H}\right)^{d+1}  \int_{0}^{\infty} \frac{\d z_1}{(z_1)^{d+1}} \int_0^{z_1} \frac{\d z_2}{(z_2)^{d+1}} \,z_2^3 z_1^3  \\[4pt]
&\ \ \ \  \times  \Big[ F'_P(z_2) F'_q(z_1)\I'_{Q_t}(z_1) F_P'(z_2)  F'_q(z_2)\K'_{Q_t}(z_2)  + \{ P \to -P \} \Big] \, , 
\end{aligned}
\label{equ:exchange1}
\eeq
where $' = z \partial_z$, $Q_t \equiv |P+q| = \sqrt{P^2 +q^2 +2 P\cdot q}$. The contribution coming from $P \to -P$ is the $u$-channel exchange, while the $t$-channel vanishes for these momenta.  Expanding in powers of $P \cdot q / q^2$, and extracting the term proportional to $P^{4\nu} q^{-d-4}$, we find 
\beq
A(q, P)_{\tilde \lambda_3^2} \supset \frac{495}{512} \tilde \lambda_3^2 \,\frac{f_\pi^4}{H^4} \frac{ (2 P\cdot q)^2 P^6}{q^7} \, ,
\eeq
where we have set $d=3$. 

\vskip 4pt
The amplitudes at order $\lambda_3^2$ and $\lambda_3 \tilde \lambda_3$ are somewhat more involved as there are now contractions of the $\lambda_3$-vertex that are not equivalent. Dropping terms suppressed by~powers of $(P/q)^2 \ll 1$, the contribution proportional to $\lambda_3^2$ is
\begin{align}\label{equ:exchange2}
A (q,P)_{\lambda_3^2}&= \frac{2}{9} \lambda_3^2   \,c_r^4\left(\frac{ f_\pi}{\tilde H}\right)^{d+1} \int_{0}^{\infty} \frac{\d z_2}{(z_2)^{d+1}} \int_0^{z_2} \frac{\d z_1}{(z_1)^{d+1}} z_1^2 z_2^2 \\
  &\ \ \  \bigg( \Big[  (P\cdot q) F_P(z_2) ( F_q(z_1)\I'_{Q_t}(z_1) + F'_q(z_1)\I_{Q_t}(z_1)) - (q^2 + P\cdot q) F'_P(z_2) F_q(z_1)\I_{Q_t}(z_1) \Big] \nonumber \\
 &\ \ \   \Big[  (P\cdot q) F_P(z_2) ( F_q(z_2)\K'_{Q_t}(z_2) + F'_q(z_2)\K_{Q_t}(z_2)) - (q^2 + P\cdot q) F'_P(z_2) F_q(z_2)\K_{Q_t}(z_2) \Big] \nonumber \\ 
  &\ \ \ + \{P \to - P\} \bigg) \nonumber \\[8pt]
 &\supset \frac{1045}{1536}  \lambda_3^2  \frac{f_\pi^4}{H^4} \frac{(2 P\cdot q)^2 P^6}{q^7} \, ,
\end{align}
where, in the final line, we have evaluated the integrals for $d=3$ and isolated the term with the correct scaling.

\vskip 4pt
The computation of the $\lambda_3 \tilde \lambda_3$ contribution is essentially a combination of the previous two.  We take the integral in~(\ref{equ:exchange2}) and turn it into two new integrals: we first replace the $z_1$-integrand in~(\ref{equ:exchange2}) with the $z_1$-integrand from~(\ref{equ:exchange1}) (times 3 for combinatorics) and then repeat the procedure, but with the $z_2$-integrand.  Performing the integrals, for $d=3$, we get 
\beq
A(q, P)_{\lambda_3 \tilde \lambda_3 }  \supset  -  \frac{605}{256}  \lambda_3 \tilde \lambda_3  \, \frac{f_\pi^4}{H^4} \frac{(2 P\cdot q)^2 P^6}{q^7} \, .
\eeq
The sign of this term is less significant than that of the other exchange diagrams, as it can be absorbed into the relative sign between the couplings $\lambda_3$ and $\tilde \lambda_3$.

\vskip 4pt
The above results are still missing terms coming from derivatives acting on the step functions that implement the $z$-ordering of the full Green's function: 
\beq
{\cal G}_Q (z_1,z_2) = \theta(z_2- z_1) \,G_Q(z_1, z_2) +  \theta(z_1- z_2) \,G_Q(z_2, z_1)\, .
\eeq
Derivatives acting on $\theta(z_i-z_j)$ produce a delta function $\delta(z_i-z_j)$, which reduces the integral to a contact interaction.  Because these terms are effectively contact terms, the internal propagator produces no additional $P\cdot q$ dependence.   
The only contributions to the spin-2 moment of the DIS amplitude can therefore arise from transverse derivatives acting on the external lines, which produces factors of $(P\cdot q)^2$. There is one such term coming from a specific contraction of the $\lambda_3^2$ diagram: 
\beq
A(q, P)_{\lambda_3^2} \supset \frac{55}{72}  \lambda_3^2 \, \frac{f_\pi^4}{H^4} \frac{(2 P\cdot q)^2 P^6}{q^7}\, .
\eeq
Had we calculated the spin-0 contribution to the amplitude, we would have had to include similar terms proportional to $\tilde \lambda_3^2$ and $\tilde \lambda_3 \lambda_3$.
 
 \vskip 10pt
Adding all the exchange contributions, we finally get
\beq
A(q, P) \supset \left[ \frac{495}{512} \tilde{\lambda}_3^2 - \frac{605}{256} \lambda_3\tilde \lambda_3 + \frac{6655}{4608}\lambda_3^2\right]  \frac{f_\pi^4}{H^4} \, \frac{(2P\cdot q)^2 P^6}{q^7} \, .
\eeq
Despite the somewhat unusual numbers appearing in the coefficients, all of these terms are needed to make the contributions from the exchange diagrams manifestly positive.  

\newpage
\section{Relation to One-Loop Anomalous Dimensions}
\label{app:betafunction}

Throughout our analysis in Section~\ref{sec:DIS}, we have used constraints on the anomalous dimension $\gamma_2$ interchangeably with the positivity of the DIS amplitude itself.  This connection appeared to arise accidentally in~(\ref{equ:spin2}) after expanding to leading order in $\gamma_2 \ll 1$.  What remains counterintuitive is that anomalous dimensions are usually associated with log-divergences and are therefore universal, while the DIS amplitude as a whole is not.  In this appendix, we will explain this apparent coincidence by demonstrating that the anomalous dimension appearing in the DIS amplitude~(\ref{equ:spin2})  necessarily agrees with the direct one-loop calculation. 

\vskip 4pt
Consider perturbing the boundary QFT by the spin-2 operator defined in \eqref{spin2def}, 
\beq
S \to S +  \lambda^{\alpha\beta} \int \d^d x \ {\cal O}^{(2)}_{\alpha\beta}\, ,
\eeq 
where $\lambda^{\alpha\beta}$ is a symmetric traceless tensor. The anomalous dimension of ${\cal O}^{(2)}_{\alpha\beta}$ can be extracted from the one-loop correction to the $\beta$-function via the relation
\beq
\label{genbetaa}
\beta^{\alpha \beta} = (d- \bar \Delta - \gamma_2)\lambda^{\alpha \beta} \, , 
\eeq
where $\bar \Delta = 2 \Delta +2$ for this spin-two operator and we have dropped terms beyond linear order in the coupling.
To determine the $\beta$-function, we calculate the power spectrum of $\O$. 
Using \eqref{equ:2pt}, the tree level contribution, at linear order in $\lambda^{\alpha \beta}$, is
\beq\label{ppptree}
\langle \O(P)\O(-P)\rangle_{\text{tree}}' \supset 2 (b_\Delta+1)A^2_{{\cal O}}  \lambda^{\alpha\beta} P_\alpha P_\beta P^{4\Delta-2d} \, ,
\eeq
where $A_\O$ and $b_\Delta$ were defined in (\ref{equ:AO}) and  (\ref{equ:bD}), respectively.
By writing the perturbation in momentum space, we see that the one-loop correction is obtained by gluing together two of the legs in the DIS amplitude,
\begin{align}
\langle \O(P)\O(-P)\rangle'_{\text{1-loop}} &= -(b_\Delta+1)\int \frac{\d^d q}{(2\pi)^d}\lambda^{\alpha\beta} q_\alpha q_\beta \,A(q, P)\\
&\supset \gamma_2  (b_\Delta+1)c(d) A_{{\cal O}}^2 P^{4\Delta-2d}  \lambda^{\alpha\beta}P^{\gamma}P^{\delta} \int \frac{\d^dq}{(2\pi)^d}\, q^{-d-4}q_\alpha q_\beta q_\gamma q_\delta \, ,
\end{align}
where, in the second line, we have substituted (\ref{equ:spin2}), as this leads to the only log-divergence relevant to the $\beta$-function. Using the symmetry of the integral and tracelessness of $\lambda^{\alpha\beta}$ to replace 
\beq
\lambda^{\alpha\beta}P^\gamma P^\delta \, q_\alpha q_\beta q_\gamma q_\delta \to \frac{2}{d(d+2)}\lambda^{\alpha\beta}P_\alpha P_\beta q^4 \ , 
\eeq
we find the following log-divergence:
\beq\label{ppploop}
 2 \gamma_2  (b_\Delta+1) A_{\cal O}^2 P^{4\Delta-2d} \lambda^{\alpha\beta}P_\alpha P_\beta \log \Lambda \, .
\eeq
Comparing \eqref{ppptree} to \eqref{ppploop}, we see that holding the power spectrum fixed requires
\beq
\beta^{\alpha \beta} = \frac{d \lambda^{\alpha \beta}}{d \log \Lambda} \supset -\gamma_2 \lambda^{\alpha\beta} \, ,
\eeq
which proves that $\gamma_2$ is indeed the one-loop anomalous dimension; cf.~\eqref{genbetaa}.

\newpage
\section{Propagation Speeds for the DBI Action}
\label{app:DBI}

In this appendix, we study the specific effective theory arising from the DBI Lagrangian:
\beq
{\cal L} = - \Lambda^4 \sqrt{1+ \frac{\partial_\mu \phi \partial^\mu \phi}{\Lambda^4} } \, .
\label{equ:DBI}
\eeq
This is the unique higher-derivative theory that is known to have a consistent UV completion. As we will see, in this case, all couplings of the EFT are determined in term of the propagation speeds $c_s$ and $c_r$, for dS and AdS, respectively. We will determine the scaling of the EFT parameters in the limit $\{c_s,c_r\} \ll 1$. 

\subsection{Inflation}

We first consider (\ref{equ:DBI}) in an inflationary background. 
We will take the flat-space limit, so that we can write $\phi = \dot \phi t + \varphi$, for $\dot \phi = const$. The Lagrangian then becomes
\bea
{\cal L} &=& - \Lambda^4 \sqrt{1- \frac{\dot\phi^2}{\Lambda^4} -\frac{2\dot \varphi \dot \phi}{\Lambda^4}+ \frac{\partial_\mu \varphi \partial^\mu \varphi}{\Lambda^4} } \\
&=& - \Lambda^4 c_s + \frac{1}{2c_s}\left[  \frac{1}{c_s^2 }\dot \varphi^2 - \partial_i \varphi \partial^i \varphi \right] +{\cal O}(\varphi^3)\, ,
\eea
where we have dropped a total derivative term proportional to $\dot \varphi$ and identified the sound speed~as
\beq
c_s^2  \equiv 1 - \frac{\dot\phi^2}{\Lambda^4}\, .
\eeq
We see that $c_s \to 0$ when $\dot \phi^2 \to \Lambda^4$.  

\vskip 4pt
To connect this to the Goldstone EFT of \S\ref{sec:EoI}, it is convenient to write the DBI action in terms of $U = t + \pi$, with  $\pi \approx \varphi /\dot \phi$.
In the limit $c_s \ll 1$, we then get
\bea
{\cal L} &\approx & - \Lambda^4 c_s  \sqrt{1 +\frac{1}{c_s^2} \left(\partial_\mu U \partial^\mu U + 1\right) }  \\
&\approx& - \Lambda^4 c_s \left[ 1+ \sum_{n=1}^{\infty}\frac{c_n}{n!}\frac{1}{c_s^{2n}}   \left(\partial_\mu U \partial^\mu U + 1\right)^n \right] ,
\eea
where $c_n = \sqrt{\pi}/ (2 \,\Gamma[3/2-n])$.  We see that DBI naturally produces $M_n \propto c_s^{-2n +1}$, as anticipated with our choice of scaling in (\ref{eq:Mncs}).

\subsection{RG Flow}

Next, we repeat the exercise for the case of slow RG flow, where the evolution of $\phi$ is in the radial direction of the AdS background.
We write $\phi =  \phi' r + \varphi$, with $\phi' = const$, so that the DBI Lagrangian becomes
\bea
{\cal L} &=& - \Lambda^4 \sqrt{1+ \frac{\phi'{}^2}{\Lambda^4} + \frac{2 \varphi' \phi'}{\Lambda^4}+ \frac{\partial_\mu \varphi \partial^\mu \varphi}{\Lambda^4} } \\[4pt]
&=& - \Lambda^4 c_r^{-1} + \frac{c_r}{2}\Big[ \dot \varphi^2 - \partial_i \varphi \partial^i \varphi - c_r^2 \varphi'{}^2 \Big] +{\cal O}(\varphi^3)\, ,
\eea
where the infall speed is 
\beq
c_r^{-2} =1+ \frac{\phi'{}^2}{\Lambda^4}\, .
\eeq
We see that $c_r \ll 1$ now requires $\phi'{}^2 \gg \Lambda^4$. 
Taking this limit, and ignoring potential concerns about the validity of the effective description,\footnote{Naively, we might expect the description to break down at the scale $\Lambda$ controlling the higher-derivative expansion in~(\ref{equ:DBI}).  However, the DBI action is UV-complete up to corrections that are proportional to the ``acceleration" of the field.  It is therefore reasonable to trust the theory even when $ \phi' \gg \Lambda^2$, provided that $\phi''$ is small.} the DBI Lagrangian becomes
\beq
{\cal L} \approx - f^4 \sqrt{1+ c_r^2 + (\partial_M U \partial^M U -1)} \, ,
\eeq
where we have defined $f^4 \equiv \Lambda^2 |\phi'|$ and $U= r+\pi$, with $\pi = \varphi / \phi'$.
Matching this to the expansion in (\ref{equ:AdSAction}), we find 
\begin{align}
M_2 &= \left(\frac{1}{4} - \frac{3}{8} c_r^2 \right) f^4\, ,  \\ 
M_3 &= \left(-\frac{3}{8}+ \frac{15}{16} c_r^2 \right)  f^4 = - \frac{3}{2}(1-c_r^2) M_2 \, .
\end{align}
In terms of the parameters $c_2$ and $c_3$ defined in (\ref{eq:Mn}), this becomes
\beq
c_3 = - \frac{3}{2} (1- c_r^2) c_2   \, .
\eeq
Using this in the sum rule (\ref{eq:sumrulefinal}), we find
\beq
(1-c_r^2) + \frac{1}{16} \frac{(1-c_r^2)^2}{c_r^2}= \frac{1}{55 \pi} \left(\frac{f_\pi}{\tilde H}\right)^4  \lim_{q \to \infty} q^3 \int_0^1 \d x\, x \,  {\rm Im}[A(x, q^2)] \, ,
\eeq
which relates the infall speed for DBI to an integral over the DIS amplitude.
Notice that for $c_r \ll 1$, the left-hand side scales as $c_r^{-2}$, which is to be compared to the scaling $c_r^{-6}$ for a generic higher-derivative theory.
This cancelation calls for a deeper explanation.

\clearpage
\phantomsection
\addcontentsline{toc}{section}{References}
\bibliographystyle{utphys}
\bibliography{Scale-Refs}

\end{document}